\documentclass[pre,aps,twocolumn,superscriptaddress]{revtex4-1}

\usepackage{graphicx}
\usepackage{amssymb,amsfonts,amsmath}
\usepackage{color}
\usepackage{ulem}
\usepackage[hidelinks]{hyperref}
\usepackage{bbold}
\usepackage{algpseudocode}
\usepackage{algorithm}

\usepackage{tikz}
\usetikzlibrary{shapes}
\usetikzlibrary{patterns}
\usetikzlibrary{angles, quotes}

\newcommand{\rv}{{\mathbf r}}

\newcommand{\Tr}{{\rm Tr}\,}
\newcommand{\e}{{\rm e}}

\newcommand{\Jv}{{\bf J}}

\newcommand{\pv}{{\bf p}}

\newcommand{\Fv}{{\bf F}}

\newcommand{\msphantom}[1]{$\ldots$}

\newcommand{\eps}{{\boldsymbol \epsilon}}

\newcommand{\eqr}[1]{Eq.~\eqref{#1}}

\newcommand{\unity}{{\mathbb 1}}

\newcommand{\mydelete}[1]{{}}
\newcommand{\taub}{{\boldsymbol\tau}}
\newcommand{\aas}{{\alpha\alpha'}}
\newcommand{\rmint}{{\rm int}}

\newcommand{\rmext}{{\rm ext}}

\newcommand{\bsig}{\boldsymbol\sigma}

\newcommand{\rmcl}{{\rm cl}}
\newcommand{\Sv}{{\bf S}}

\newcommand{\rmi}{{\rm i}}

\newcommand{\calU}{{\cal U}}
\newcommand{\Cv}{{\bf C}}
\newcommand{\alphas}{{\alpha'}}
\renewcommand{\rho}{n}

\newcommand{\ii}{{\rm i}}
\newcommand{\ccdot}{\, \cdot \,}

\newcommand{\Hcal}{\mathcal{\hat H}}
\newcommand{\Acal}{\mathcal{A}}
\newcommand{\Jcal}{\mathcal{J}}
\newcommand{\Ocal}{\mathcal{O}}
\newcommand{\Pcal}{\mathcal{P}}

\renewcommand{\b}[1]{\mathbf{#1}}
\newcommand{\bg}[1]{\boldsymbol{#1}}
\newcommand{\bhat}[1]{\hat {\mathbf{#1}}}
\newcommand{\bilde}[1]{\tilde {\mathbf{#1}}}
\newcommand{\bildehat}[1]{\tilde {\hat {\mathbf{#1}}}}
\newcommand{\Hhat}{\hat H}

\DeclareMathOperator{\sgn}{\mathrm{sgn}}
\DeclareMathOperator{\cov}{\mathrm{cov}}

\begin{document}

\title{Quantum statistical mechanics: Gauge invariance, operator shifting,\\
  hyperdensity functionals, and nonequilibrium sum rules}

\author{Johanna M\"uller}
\affiliation{Theoretische Physik II, Physikalisches Institut, 
  Universit{\"a}t Bayreuth, D-95447 Bayreuth, Germany}

\author{Matthias Schmidt}
\affiliation{Theoretische Physik II, Physikalisches Institut, 
  Universit{\"a}t Bayreuth, D-95447 Bayreuth, Germany}
\email{Matthias.Schmidt@uni-bayreuth.de}

\date{24 September 2025, revised version: 22 July 2026}

\begin{abstract}
   We provide an extended account of the recent statistical mechanical
   theory of gauge invariance against operator shifting in quantum
   many-body systems (\href{https://doi.org/10.48550/arXiv.2509.20494}
   {arXiv:2509.20494}). The gauge transformation is enacted by a
   shifting superoperator that displaces the fundamental position and
   momentum degrees of freedom. The shifting superoperator constitutes
   a map between Hilbert space operators and it features Lie algebra
   commutator structure. Averages of general observables remain
   invariant under the shifting both in and out of thermal
   equilibrium, as well as in groundstates. The gauge invariance
   induces exact sum rules that interconnect global observables and
   associated locally resolved correlation functions.  In particular
   we describe the resulting one-body force, hyperforce, product, and
   two-body sum rules. We relate the shifting superoperator to a
   previously formulated quantum canonical transformation and present
   the generalization of quantum shifting to multi-component systems.
   The gauge theory respects fundamental fermionic and bosonic
   particle properties, as we demonstrate by proving the compatibility
   of operator shifting and exchange symmetry.  We formulate the
   quantum version of hyperdensity functional theory to provide formal
   access to hyperforces as well as to general averaged quantum
   observables via universal density functionals.  For time-dependent
   situations, we describe quantum dynamical gauge invariance and
   prove exact dynamical sum rules for nonequilibrium situations, as
   generated by Hamiltonian time dependence.  We argue for the
   fundamental status of statistical mechanical gauge invariance based
   on the compliance of the underlying geometry with canonical
   quantization according to Dirac's correspondence principle.
   Analogies and differences of the quantum mechanical sum rules with
   their classical counterparts remain indicative of the respective
   levels of description.
\end{abstract}

\maketitle

\section{Introduction}
\label{SECintroduction}

Gauge invariance constitutes a versatile construction principle for
the development of fundamental physical theories. The recent
application of gauge invariance to the quantum statistical mechanics
of many-body systems~{\cite{mueller2025quantum, hermann2022quantum,
  phamvan2026quantumGeometry, phamvan2026exchangePrivateLindblad}}
provides a topical example for much wider efforts \cite{jackson2001,
  raifeartaigh2000}.  The birth of the gauge concept is often
attributed to Emmy Noether's 1918 landmark publication on invariant
variational problems \cite{noether1918}, but the history of the
development of her ideas is complex \cite{brewer1981,
  neuenschwander2011} and the origins of electrodynamics, as arguably
the most iconic physical gauge theory~\cite{jackson2001,
  raifeartaigh2000}, are much older than Noether's mathematical
insights into the general problem.

The gauge invariance of {\it classical} statistical mechanics against
a specific shifting transformation was explored in a range of recent
contributions, including spatially homogeneous
\cite{hermann2021noether, hermann2022topicalReview,
  hermann2022variance} and inhomogeneous \cite{tschopp2022forceDFT,
  sammueller2022forceDFT, sammueller2023whatIsLiquid,
  hermann2023whatIsLiquid, robitschko2024any} shifting. The framework
leads to novel force and `hyperforce' sum rules that are exact and
that have been demonstrated to be accessible in numerical simulation
work \cite{sammueller2023whatIsLiquid, hermann2023whatIsLiquid,
  robitschko2024any}. Performing such simulations is arguably more
straightforward for classical systems than it is for quantum
particles.  The classical hyperforce concept is intimately related to
a corresponding hyperdensity functional formulation for the
equilibrium behaviour of general observables
\cite{sammueller2024hyperDFT, sammueller2024whyhyperDFT,
  sammueller2024multihyperDFT}. These target observables of interest
are referred to as `hyperobservables' in the present context
\cite{sammueller2024hyperDFT, sammueller2024whyhyperDFT,
  sammueller2024multihyperDFT, mueller2024gauge, mueller2024whygauge,
  mueller2024dynamic, matthes2024mix}, where the terminology draws on
Hirschfelder's use of the prefix in a different context to express
general applicability \cite{hirschfelder1960}.

That phase space shifting \cite{hermann2021noether,
  hermann2022topicalReview, hermann2022variance, hermann2022quantum,
  tschopp2022forceDFT, sammueller2022forceDFT,
  sammueller2023whatIsLiquid, hermann2023whatIsLiquid,
  robitschko2024any} constitutes a nontrivial gauge transformation of
statistical mechanics is a recent insight \cite{mueller2024gauge,
  mueller2024whygauge, mueller2024dynamic, matthes2024mix}. Popular
accounts have been given~\cite{rotenberg2024spotted,
  miller2025physicsToday} and significant extensions of the concept
include the rotational theory for anisotropic bodies by
Nguyen-Tran-Thanh {\it et al.}~\cite{nguyen2026}, the functional
analysis formulation based on the Schwartz space and its dual by
Maruyama {\it et al.}~\cite{maruyama2026}, and Pham-Van's
investigation of the interrelationships with standard symmetries
\cite{phamvan2026symmetry}.  The hyperforce sum rules connect to the
body of exact statistical mechanical sum rules for the liquid state
that were developed in classical work, such as that by Baus
\cite{baus1984}, Henderson \cite{henderson1992},
Evans~\cite{evans1979}, Mikheev and Weeks \cite{mikheev1991}, and Di
Caprio {\it et al.}~\cite{dicaprio1996}.  The latter
authors~\cite{dicaprio1996} have pointed out analogies to the
important Takahashi-Ward identities of quantum field theory
\cite{takahashi1957, ward1950}, as recognized in
Ref.~\cite{sammueller2023whatIsLiquid} and much expanded on in
Ref.~\cite{nguyen2026}.

The shifting invariance of {\it quantum} statistical mechanics was
demonstrated for a specific operator shifting
map~\cite{hermann2022quantum} via the application of the general
concept of quantum canonical transformations \cite{anderson1994}. This
method constitutes a proof of the exact spatially localized force
density balance, which is relevant in several recent force-related
quantum theoretical approaches \cite{tokatly2005one, tokatly2005two,
  tokatly2007, ullrich2006, tchenkoue2019, tarantino2021}, see the
brief overview given in Ref.~\cite{hermann2022quantum}.  {We commment
  on Pham-Van's very recent work \cite{ phamvan2026quantumGeometry,
    phamvan2026exchangePrivateLindblad} in the concluding section.}

Interacting quantum particles generate rich and diverse physics with a
broad range of interesting emergent phenomena and effects. In
particular machine-learning methods offer great potential for making
much headway, see Ref.~\cite{carleo2017} for early pioneering work and
Ref.~\cite{huang2023review} for a density functional perspective.  The
specific realm of quantum many-body dynamics of ultracold atomic
systems constitutes a highly active research area \cite{bloch2008}
with seminal studies addressing many-body localization in two
dimensional systems both experimentally \cite{choi2016, bordia2017}
and theoretically \cite{yan2017prl,yan2017pra}. We comment on the
potential applicability of our framework to these and further relevant
quantum systems below.

Here we give an extended account of the quantum gauge invariance
theory of Ref.~\cite{mueller2025quantum}. Our presentation constitutes
a standalone formulation of the quantum physics, which demonstrates
that the theory does {\it not} rely on the corresponding classical
framework, but rather that the shifting gauge invariance constitutes
an intrinsic property of quantum statistical mechanics.  We lay out
the connections with the quantum canonical transformation of
Ref.~\cite{hermann2022quantum} and describe similarities and important
differences with the classical theory based on Dirac's correspondence
principle.

\begin{figure}[tb]
  \includegraphics[page=1,width=.82\columnwidth]{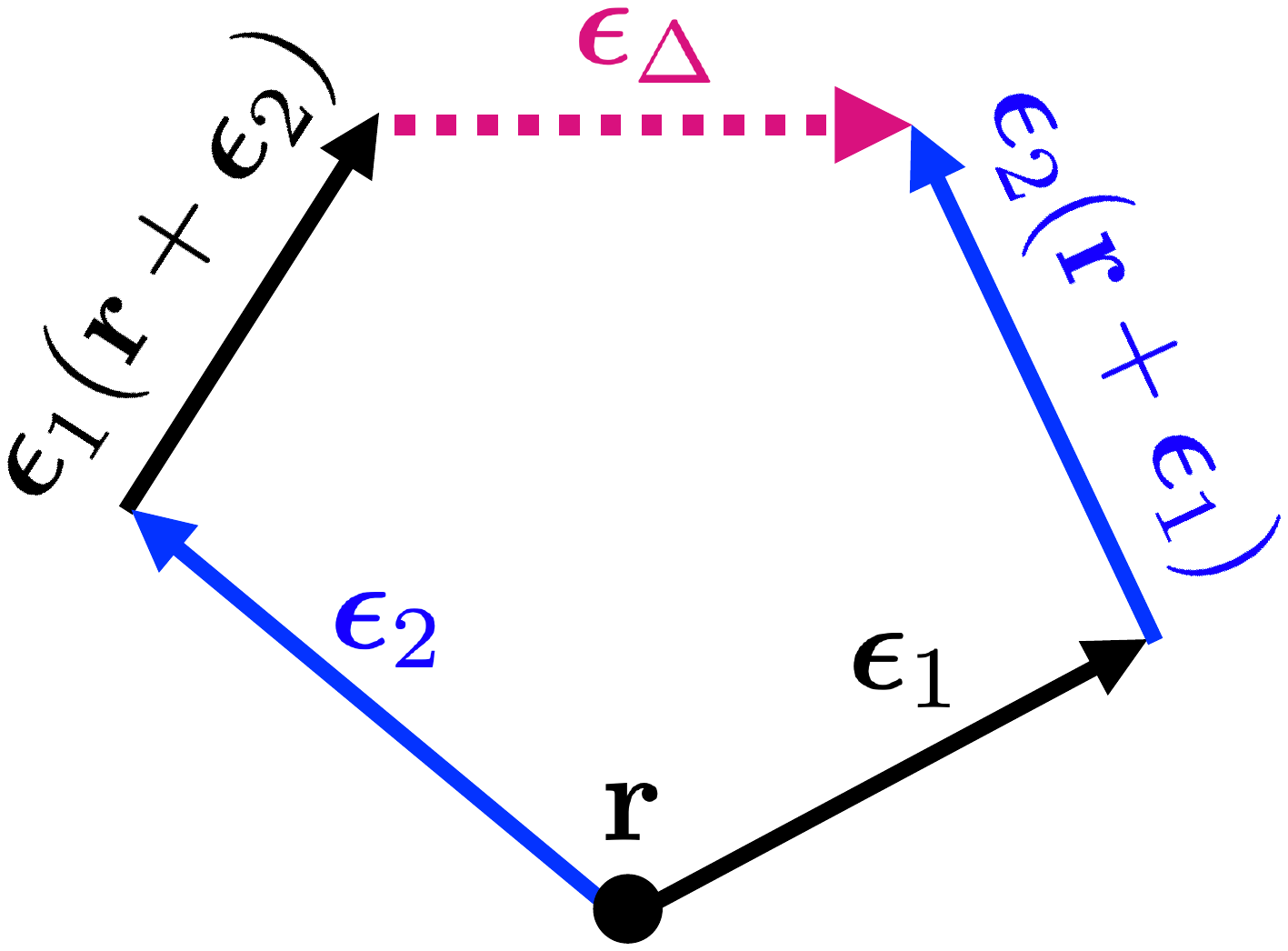}
  \caption{Illustration of the geometry that underlies the Lie
    algebra of quantum operator shifting. One considers two
    consecutive shifts by vector fields $\eps_1(\rv)$ and
    $\eps_2(\rv)$, where $\rv$ is the spatial coordinate.  The finite
    difference in displacement that follows from exchanging the order
    of the two shifts is $\eps_1(\rv) + \eps_2(\rv+\eps_1(\rv)) -
    \eps_2(\rv) - \eps_1(\rv+\eps_2(\rv))$. Expanding to lowest order
    in the field gradients gives the Lie bracket of the two vector
    fields, $\eps_\Delta(\rv)=\eps_1(\rv) \cdot \nabla \eps_2(\rv) -
    \eps_2(\rv) \cdot \nabla \eps_1(\rv)$, where~$\nabla$ indicates
    the derivative with respect to $\rv$. In the statistical
    mechanical description the vector fields are evaluated at position
    $\rv_i$ of particle $i$. Noncommutativity occurs already
    classically and it persists under quantization.  The momentum
    operators are transformed consistently and the joint
    transformation leaves all statistical mechanical averages
    invariant, both classically \cite{mueller2024gauge,
      mueller2024whygauge, mueller2024dynamic, matthes2024mix,
      rotenberg2024spotted, miller2025physicsToday, nguyen2026,
      maruyama2026, phamvan2026symmetry} and quantum mechanically
    \cite{mueller2025quantum, hermann2022quantum,
      phamvan2026quantumGeometry, phamvan2026exchangePrivateLindblad}.
    The quantum shifting superoperator algebra is described in
    Sec.~\ref{SECequilibriumShifting}, see the Lie bracket
    \eqref{EQepsDelta}, and proven in the
    Appendices~\ref{APPderivationLie} and
    \ref{APPcommutatorShiftingOperators}. An in-depth description of
    the geometrical group structure, including inversion and
    associativity, is provided in Ref.~\cite{mueller2024whygauge} and
    the (classical) relationship with standard symmetries is developed
    in Ref.~\cite{phamvan2026symmetry}.}
  \label{FIG1}
\end{figure}

{ 

The quantum operator shifting constitutes a continuous (Lie) group,
where each group element is parameterized by an associated shifting
vector field \cite{mueller2025quantum}. The generators of the shifting
group follow as the infinitesimal versions of these in general finite
transformations.  The generators are essential in epitomizing the
fundamental mathematical structure by forming a Lie algebra. This
implies that the commutator of two consecutive infinitesimal shifts is
again an element of the Lie algebra. An important consequence is an
implied `closed' algebraic structure, such that no uncontrolled
proliferation of potentially ever more complex transformations
occurs. 

As an essential concept, the quantum shifting is enacted by shifting
superoperators, i.e.\ mappings between Hilbert space operators, and
these induce exact sum rules that apply to the quantum statistical
mechanics.  The sum rules form a controlled and again closed set of
constraints for measurable correlation and response functions.
Specifically, the algebraic Lie structure is expressed via the Lie
bracket of the two shifting vector fields, which is a fundamental
geometrical structure that maps two given `input' vector fields to a
specific `output' vector field.  An illustration of the geometric
nature of the Lie bracket is given in Fig.~\ref{FIG1} and we refer the
Reader to Ref.~\cite{mueller2024whygauge} for a detailed presentation
of the classical case, of which the salient features remain applicable
in the present quantum treatment, as we demonstrate.

}

{
For numerical applications of our framework to concrete systems, we
refer the Reader to the simplistic quantum harmonic oscillator toy
model \cite{mueller2025quantum}, as well as to the much more
substantial investigations by Pham-Van based on numerically exact
diagonalization techniques \cite{phamvan2026quantumGeometry,
  phamvan2026exchangePrivateLindblad}.
We also note the body of {\it classical} statistical gauge geometry
simulation studies, which range from investigations of the fluid pair
force correlation structure \cite{sammueller2023whatIsLiquid}, to the
fluid adsorption behaviour at planar walls \cite{robitschko2024any,
  matthes2024mix}, and the dynamical force structure in complex
many-body systems \cite{mueller2024dynamic}.

}

This paper is organized as follows.
We describe the considered quantum statistical mechanics in
Sec.~\ref{SECquantumStatMech}, beginning with the
microscopic quantum many-body Hamiltonian and its
equilibrium statistical mechanics in
Sec.~\ref{SECquantumLongManyBodyModel}.
The standard one-body observables, including density, current, and
force density operators, are introduced in
Sec.~\ref{SECquantumLongOneBodyObservables}.
One-body hyperforce and hyperfluctuation profiles are defined in
Sec.~\ref{SECquantumLongHyperforceDefinition}.
We start the description of the equilibrium gauge invariance in
Sec.~\ref{SECequilibriumGaugeInvariance} by giving an abridged account
of the results presented in Ref.~\cite{mueller2025quantum}, which
includes the equilibrium operator shifting in
Sec.~\ref{SECequilibriumShifting},
the derivation of force and hyperforce sum rules 
in Sec.~\ref{SECequilibriumSumRules},
as well as product and two-body hyperforce sum rules
in Sec.~\ref{SEChigherSumRules}.
We describe key aspects of the operator shifting in
Sec.~\ref{SECaspects}.
The relationship with the quantum canonical shifting
transformation~\cite{hermann2022quantum} is laid out in
Sec.~\ref{SECquantumCanonicalTransformation}.
The compatibility of particle exchange symmetry with the gauge
shifting is demonstrated in Sec.~\ref{SECexchangeSymmetryAndShifting}.
The generalization of operator shifting to multi-component systems is
given in Sec.~\ref{SECmulticomponent}.

We develop the quantum hyperdensity functional theory in
Sec.~\ref{SEChyperDensityFunctionalTheory}, where we describe a
suitably extended ensemble in Sec.~\ref{SECextendedEnsemble}, its
corresponding density functional minimization principle in
Sec.~\ref{SECquantumLongMinimizationPrinciple}, the
hyper-Ornstein-Zernike relation in Sec.~\ref{SECquantumLongHyperOZ},
and the unique way to express general observables as density
functionals in Sec.~\ref{SECquantumLongHyperdensityFunctionals}.
Dynamical gauge invariance is described in
Sec.~\ref{SECdynamicalGaugeInvariance}, starting with a definition of
the nonequilibrium many-body setup in
Sec.~\ref{SECquantumLongManyBodyDynamics}, which is followed by the
demonstration of the consequences of dynamical shifting in
Sec.~\ref{SECquantumLongDynamicalOperatorShifting}.
We use Dirac's correspondence principle to lay out the relationship of
the quantum theory with the corresponding classical version in
Sec.~\ref{SECquantumLongDiracCorrespondence}, starting from a
description of the classical many-body statistical mechanics in
Sec.~\ref{SECquantumLongClassicalManyBodyDescription} and then
relating the respective mathematical formulations of the shifting
transformation to each other in
Sec.~\ref{SECquantumLongClassicalShiftingAndQuantumShifting}.
We give our conclusions in Sec.~\ref{SECquantumLongConclusions}, where
we also lay out the structure of the Appendices that contain important
detailed derivations of all key results that are presented in the main
text.

{

The Letter~\cite{mueller2025quantum} presented the equilibrium
operator shifting (Sec.~\ref{SECequilibriumShifting}), the force and
hyperforce sum rules (Sec.~\ref{SECequilibriumSumRules}), and the
connection of the superoperator formalism with the quantum canonical
shifting transformation
(Sec.~\ref{SECquantumCanonicalTransformation}). All further quantum
results that we present here have not appeared elsewhere, including
the product and two-body equilibrium hyperforce sum rules
(Sec.~\ref{SEChigherSumRules}), the considerations of particle
exchange symmetry (Sec.~\ref{SECexchangeSymmetryAndShifting}), the
treatment of multi-component systems (Sec.~\ref{SECmulticomponent}),
the quantum hyperdensity functional theory
(Sec.~\ref{SEChyperDensityFunctionalTheory}), the quantum
hyper-Ornstein-Zernike relation (Sec.~\ref{SECquantumLongHyperOZ}),
the quantum dynamical gauge invariance
(Sec.~\ref{SECdynamicalGaugeInvariance}), which generalizes the recent
classical dynamical treatment \cite{mueller2024dynamic}, and the
relationship of the present quantum theory with the corresponding
classical statistical mechanical gauge theory
(Sec.~\ref{SECquantumLongDiracCorrespondence}).}

\section{Quantum statistical mechanics}
\label{SECquantumStatMech}

We first describe the required standard ingredients for the
development of the quantum shifting gauge theory. Our purpose is to
clarify notation and to spell out explicitly the involved quantum
statistical mechanical concepts, including the many-body Hamiltonian
(Sec.~\ref{SECquantumLongManyBodyModel}), the relevant one-body
observables (Sec.~\ref{SECquantumLongOneBodyObservables}), and the
hyperforce and hyperfluctuation profiles that describe the correlation
behaviour of general observables
(Sec.~\ref{SECquantumLongHyperforceDefinition}).  { Our
  presentation is consistent with and more detailed than what we
  provide in the Letter \cite{mueller2025quantum}.}

\subsection{Microscopic many-body model}
\label{SECquantumLongManyBodyModel}

We consider systems that are described by Hamiltonians that possess
the following standard form:
\begin{align}
   \Hhat &= \sum_i \frac{\hat \pv_i^2}{2m}
  + u(\rv^N) + \sum_i V_\rmext(\rv_i),
  \label{EQHamiltonian}
\end{align}
where the sums run over all particles $i=1,\ldots, N$, with total
number of particles $N$, the variable $m$ indicates the particle mass,
$\hat\pv_i = -\rmi\hbar \nabla_i$ is the momentum operator of particle
$i$, the roman letter $\rmi$ denotes the imaginary unit, $\hbar$ is
the reduced Planck constant, $\nabla_i$ indicates the derivative with
respect to position $\rv_i$, the interparticle interaction potential
$u(\rv^N)$ depends on all particle position coordinates
$\rv^N=\rv_1,\ldots,\rv_N$, and $V_\rmext(\rv)$ is the external
potential, here expressed as a function of the generic position
variable~$\rv$.

The microscopic position and momentum degrees of freedom satisfy the
standard canonical commutator relationships between positions,
$[\rv_i, \rv_j]=0$, between momenta, $[\hat\pv_i, \hat\pv_j]=0$, and
between mixed pairs, $[\rv_i, \hat\pv_j] = \rmi \hbar \delta_{ij}
\unity$, where $\delta_{ij}$ is the Kronecker symbol, $\unity$ denotes
the $d\times d$ unit matrix, and $d$ indicates the spatial
dimensionality. The commutator of two general Hilbert space operators
$\hat A$ and $\hat B$ is defined in the standard way as $[\hat A, \hat
  B]= \hat A\hat B - \hat B \hat A$.

We consider thermal equilibrium and hence let the Hamiltonian carry no
explicit dependence on time, as indicated by the subscript zero in
$\Hhat_0$. The corresponding canonical partition sum is $Z=\Tr \e^{-\beta
  \Hhat_0}$, where $\beta = 1/(k_BT)$ with Boltzmann constant $k_B$ and
absolute temperature $T$.  The canonical trace is defined as
$\Tr\,\cdot\, = \sum_n \langle n|\,\cdot\,|n\rangle$, where
$|n\rangle$ denotes a general basis of the $N$-body Hilbert space in
Dirac's bracket notation \cite{dirac1958book, dirac1939notation}.
Whether the quantum particles are fermions or bosons is set by the
exchange symmetry that the Hilbert space basis~$|n\rangle$ satisfies;
we return to this point in the context of operator shifting below in
Sec.~\ref{SECexchangeSymmetryAndShifting}.  From the partition
function $Z$ the canonical free energy follows in the standard way as
$F=-k_BT \ln Z$ and equilibrium averages in the canonical ensemble are
given by $\langle\,\cdot\,\rangle = \Tr \,\cdot\, \e^{-\beta \Hhat_0}/Z$.

The statistical mechanical treatment below continues to hold in the
grand ensemble, where the particle number~$N$ is a fluctuating
quantity and the chemical potential $\mu$ is used instead as the
prescribed control parameter. We defer the detailed specification of
the grand ensemble to Sec.~\ref{SEChyperDensityFunctionalTheory},
where we will use it as the basis of the hyperdensity functional
formulation.

The Mori product (or `Mori-Kubo-Bogoliubov' product)~is a common means
to express correlation and response functions and it constitutes
formally an inner product between two operators
\cite{fick1990book}. The Mori product of the Hilbert space operators
$\hat A$ and $\hat B$ is defined by
\begin{align}
  (\hat A | \hat B) &=
  \beta^{-1} \int_0^\beta d\beta' \Tr \hat A^\dagger \e^{-\beta'\Hhat_0}
  \hat B \e^{\beta'\Hhat_0} \frac{\e^{-\beta \Hhat_0}}{Z},
  \label{EQlongQuantumMoriProductDefinintion1}
\end{align}
where the operator $\hat A^\dagger$ is the adjoint operator of $\hat
A$. An alternative definition that is equivalent to
\eqr{EQlongQuantumMoriProductDefinintion1} is based on the Heisenberg
picture and given by~\cite{sauermann1996}
\begin{align}
  (\hat A|\hat B) &=
  \beta^{-1}\int_0^\beta d\beta' \langle \hat A^\dagger \hat B({\rm
  i}\hbar \beta')\rangle,
\end{align}
where $\hat B(t)$ denotes the Heisenberg operator that corresponds to
the Schr\"odinger operator $\hat B$, and the Heisenberg operator is
evaluated at the imaginary time $t={\rm i}\hbar \beta'$.  We give
further background on the Mori product in
Appendix~\ref{APPinnerProducts} and also refer the Reader to
Ref.~\cite{fick1990book} for a pedagogical account. The Mori product
is linear in its second argument and it is complex conjugate linear in
its first argument.  In general its value is complex and the Mori
product is conjugate symmetric with respect to an exchange of its two
arguments, $(\hat A | \hat B) = (\hat B | \hat A)^\star$, where the
asterisk denotes complex conjugation. In the case of physical quantum
observables, i.e.\ self-adjoint operators $\hat A=\hat A^\dagger$ and
$\hat B = \hat B^\dagger$, the Mori product is real, $(\hat A|\hat
B)\in \mathbb{R}$. Proofs of these standard properties
\cite{fick1990book} are given in Appendix~\ref{APPinnerProducts} for
completeness.

It is furthermore useful to also define the corresponding Mori
covariance of two general operators $\hat A$ and $\hat B$ as
\cite{fick1990book}
\begin{align}
  {\rm cov}(\hat A|\,\hat B) &=
  (\hat A|\hat B) - \langle \hat A^\dagger \rangle \langle \hat B\rangle, 
  \label{EQlongQuantumMoriCovariance}
\end{align}
where as before $(\hat A|\hat B)$ indicates the Mori product
\eqref{EQlongQuantumMoriProductDefinintion1} and the angular brackets
denote the thermal equilibrium average as defined above.

\subsection{Standard one-body observables}
\label{SECquantumLongOneBodyObservables}

When aiming at a reduced statistical mechanical description of the
many-body physics, it is standard methodology to work with
position-resolved quantities.  Such relevant one-body observables are
the position-resolved density operator $\hat\rho(\rv)$ and the
(scaled) current density operator $m\hat\Jv(\rv)$, which are
respectively given by
\begin{align}
  \hat\rho(\rv) &= \sum_i \delta(\rv-\rv_i),
  \label{EQlongQuantumDensityOperator}\\
  m\hat\Jv(\rv) &= \frac{1}{2}  \sum_i 
  \big[\hat\pv_i\delta(\rv-\rv_i) + \delta(\rv-\rv_i)\hat\pv_i \big].
  \label{EQlongQuantumCurrentOperator}
\end{align}
The one-body density and current density operators are connected
dynamically by the continuity equation, see
e.g.~Ref.~\cite{schmidt2022rmp}. Of further interest is the (total)
force density operator~$\hat\Fv(\rv)$, which consists of the following
sum of three contributions:
\begin{align}
  \hat\Fv(\rv) &= \nabla\cdot\hat\taub(\rv) 
  + \hat\Fv_\rmint(\rv) -\hat\rho(\rv) \nabla V_\rmext(\rv),
  \label{EQlongQuantumForceDensityOperator}
\end{align}
where the product $-\hat\rho(\rv)\nabla
V_\rmext(\rv)$ constitutes the external force density that arises from
the external force field $-\nabla V_\rmext(\rv)$ with $\nabla$
indicating the derivative with respect to position $\rv$.

The position-resolved kinetic stress density operator $\hat
\taub(\rv)$ is given by~\cite{schmidt2022rmp}:
\begin{align}
  \hat \taub(\rv) &=
  \frac{\hbar^2}{4m} \nabla\nabla \hat\rho(\rv)
  -\sum_i \frac{\hat\pv_i\delta(\rv-\rv_i)\hat\pv_i
    + \hat\pv_i\delta(\rv-\rv_i)\hat\pv_i^{\sf T}}{2m},
  \label{EQlongQuantumKineticStressOperator}
\end{align}
where $\nabla\nabla\hat\rho(\rv)$ is the spatial Hessian of the
density operator, the pairs of momentum operators form dyadic
products, and the superscript ${\sf T}$ indicates the transposition of
the corresponding $d\times d$ matrix.

The localized interparticle force density operator $\hat
\Fv_\rmint(\rv)$ is given by
\begin{align}
  \hat \Fv_\rmint(\rv) &=
  -\sum_i \delta(\rv-\rv_i)[\nabla_i u(\rv^N)],
  \label{EQlongQuantumFintOperator}
\end{align}
where the square brackets on the right hand side limit the scope of
the derivative $\nabla_i$, such that the vector field $[\nabla_i
  u(\rv^N)]$ operates merely by multiplication on the $N$-body Hilbert
space.  The force density operator \eqref{EQlongQuantumFintOperator}
is connected to the time derivative of the current density operator
\eqref{EQlongQuantumCurrentOperator} by the one-body Heisenberg
equation of motion, see again e.g.\ Ref.~\cite{schmidt2022rmp} for a
detailed description.

\subsection{Hyperforce and hyperfluctuation profiles}
\label{SECquantumLongHyperforceDefinition}

To generalize beyond the above `mechanical' one-body observables
\eqref{EQlongQuantumDensityOperator}--\eqref{EQlongQuantumFintOperator},
we also consider a general observable $\hat A(\rv^N, \hat\pv^N)$ that
shall be of interest for a given system; here
$\hat\pv^N=\hat\pv_1,\ldots,\hat\pv_N$ denotes the collection of all
momentum operators.  It is useful to refer to the role that $\hat A$
plays in the present context as a `hyperobservable' to express the
special status that this operator assumes in the framework. Thereby
$\hat A$ can constitute a global object that does not carry dependence
on a generic position variable $\rv$ or, alternatively, $\hat A$ can
be an operator that depends on $\rv$ and thus generates a correlation
function upon averaging. Hence the one-body operators
\eqref{EQlongQuantumDensityOperator} and
\eqref{EQlongQuantumCurrentOperator} are examples thereof, but $\hat
A$ can feature more general dependence on both $\rv^N$ and
$\hat\pv^N$.

For any thus given form of $\hat A$ an associated hyperforce density
observable $\hat\Sv_A(\rv)$ is defined in generalization
\cite{mueller2025quantum} of the corresponding classical phase space
function \cite{robitschko2024any, mueller2024gauge,
  mueller2024whygauge}.  For the special case of position-only
dependent hyperobservables $\hat A(\rv^N)$, this is given explicitly
by~\cite{mueller2025quantum}
\begin{align}
  \hat \Sv_A(\rv) 
  &= \sum_i \delta(\rv-\rv_i) [\nabla_i\hat A(\rv^N)].
  \label{EQlongQuantumSAOperatorPositions}
\end{align}
The form \eqref{EQlongQuantumSAOperatorPositions} is in formal analogy
to the interparticle force density observable $\hat\Fv_\rmint(\rv)$
defined in \eqr{EQlongQuantumFintOperator} upon identifying $\hat A =
-u(\rv^N)$; the sign convention is chosen to simplify subsequent
manipulations.

Given a general observable $\hat A$ that may carry dependence on both
positions $\rv^N$ and momenta $\hat\pv^N$, the corresponding
hyperforce density operator is defined~\cite{mueller2025quantum} as
the following commutator:
\begin{align}
  \hat \Sv_A(\rv) 
  &= -\frac{\rmi}{\hbar}[\hat A, m\hat\Jv(\rv)],
  \label{EQlongQuantumSAOperatorGeneral}
\end{align}
where we recall the scaled current density operator $m\hat\Jv(\rv)$
having the standard definition \eqref{EQlongQuantumDensityOperator}.
If $\hat A$ is self-adjoint, then $\hat \Sv_A(\rv)$ is also
self-adjoint and hence it constitutes a physical
observable~\cite{mueller2025quantum}. It is straightforward to show
that the general form \eqref{EQlongQuantumSAOperatorGeneral} reduces
to the gradient form \eqref{EQlongQuantumSAOperatorPositions} when
$\hat A$ is independent of the momenta.  The associated hyperforce
density distribution is the thermal average
\begin{align}
  \Sv_A(\rv) &= \langle \hat \Sv_A(\rv) \rangle.
  \label{EQlongQuantumSVAprofile} 
\end{align}

Given a specific form of the hyperobservable $\hat A$ an associated
`hyperfluctuation' profile $\chi_A(\rv)$ provides a measure of the
spatially localized behaviour of fluctuations. The hyperfluctuation
profile is defined as
\begin{align}
  \chi_A(\rv) &= {\rm cov}\big(\hat A | \hat\rho(\rv) \big),
  \label{EQhyperFluctuationProfile}
\end{align}
where ${\rm cov}(\,\cdot\,|\,\cdot\,)$ denotes the Mori covariance
\eqref{EQlongQuantumMoriCovariance} and we recall the density operator
$\hat\rho(\rv)$ in its standard
form~\eqref{EQlongQuantumDensityOperator}.  We show in the following
that the hyperforce density \eqref{EQlongQuantumSVAprofile} and
hyperfluctuation profile~\eqref{EQhyperFluctuationProfile} arise
naturally when considering the statistical mechanical gauge
invariance.

\section{Equilibrium gauge invariance}
\label{SECequilibriumGaugeInvariance}

We first give an abridged account of the results presented in
Ref.~\cite{mueller2025quantum}, beginning with a description of the
operator shifting for quantum systems in thermal equilibrium
(Sec.~\ref{SECequilibriumShifting}) and turning to the derivation of
the resulting thermal force and hyperforce sum rules
(Sec.~\ref{SECequilibriumSumRules}). We then present product and
two-body hyperforce sum rules (Sec.~\ref{SEChigherSumRules}) that go
beyond those of Ref.~\cite{mueller2025quantum}.

\subsection{Shifting superoperator algebra}
\label{SECequilibriumShifting}

The gauge theory works on the level of maps between Hilbert space
operators, which in general are referred to as `superoperators'
\cite{fick1990book}.
The `shifting' of Hilbert space operators \cite{mueller2025quantum} is
represented by two closely related types of superoperators. One is the
spatially localized vectorial shifting superoperator
$\bsig(\rv)$. This version is useful for the construction of concrete
sum rules that interconnect different types of correlation
functions. An alternative and closely related version is the
integrated scalar shifting superoperator $\Sigma[\eps]$, which depends
functionally on the given shifting field $\eps(\rv)$. The integrated
shifting superoperator enables one to very clearly formulate the
geometric relationship of the superoperator algebra with the standard
geometric Lie bracket of two vector fields. The spatially localized
and the integrated version are trivially related to each other by
multiplication with the shifting field, akin to multiplication by a
test function, and spatial integration. This operation is reversed by
functional differentiation with respect to the shifting field. We give
a compact account in the following and refer the Reader to
Ref.~\cite{mueller2025quantum} for a complete description.

Specifically, the spatially localized shifting superoperator is
defined via the (scaled) commutator with the one-body current density
operator \eqref{EQlongQuantumCurrentOperator} as
\begin{align}
  \bsig(\rv) &= -\frac{\rmi}{\hbar}[\,\cdot\,,m\hat \Jv(\rv)].
  \label{EQqsigDefinition}
\end{align}
Clearly, when applying \eqr{EQqsigDefinition} to a Hilbert space
operator the result is again a Hilbert space operator
\cite{mueller2025quantum} and thus $\bsig(\rv)$ constitutes a quantum
`superoperator' \cite{fick1990book}. We defer a discussion of the
relationship with classical Poisson brackets and Dirac's
correspondence principle to
Sec.~\ref{SECquantumLongDiracCorrespondence}.

We list three key properties of $\bsig(\rv)$. First, when applied to
the identity operator, trivially one obtains
\begin{align}
  \bsig(\rv) 1 &= 0,
  \label{EQqsigIdentiy}
\end{align}
as follows from the definition~\eqref{EQqsigDefinition} and will be
important for the derivation of the force density balance below.

The localized shifting superoperator is anti-self-adjoint:
\begin{align}
  \bsig^\dagger(\rv) &= -\bsig(\rv).
  \label{EQqsigAntiSelfAdjoint}
\end{align}
The concept of adjoining a given superoperator generalizes the
standard operation of adjoining a Hilbert space operator. Briefly,
given a superoperator~$\cal O$, its adjoint ${\cal O}^\dagger$ has the
property $\Tr \hat A^\dagger {\cal O} \hat B= \Tr ({\cal O}^\dagger
\hat A)^\dagger \hat B$, for general Hilbert space operators $\hat A$
and $\hat B$.  The standard adjoining of Hilbert space operators,
based on the Hilbert-Schmidt inner product as described in
Appendix~\ref{APPinnerProducts}, serves as background for introducing
the concept of adjoining of superoperators, see
Appendix~\ref{APPinvarianceTrace}.

The commutator of two localized shifting superoperators is
\begin{align}
  [\bsig(\rv),\bsig(\rv')] &= 
  [\nabla \delta(\rv-\rv')]\bsig(\rv)
  +\bsig(\rv')[\nabla \delta(\rv-\rv')],
  \label{EQqsigCommutator}
\end{align}
where $\rv,\rv'$ are two position variables, as before
$\delta(\,\cdot\,)$ indicates the Dirac distribution in $d$
dimensions, and the square brackets around the gradient of the Dirac
distributions limit the scope of the spatial derivative operator such
that these (bracketed) terms act on Hilbert space elements by mere
multiplication. We give in
Appendix~\ref{APPcommutatorShiftingOperators} the proof of the
commutator relationship~\eqref{EQqsigCommutator}, which carries
significance due to the fact that the quantum degrees of the $N$
particles are entirely contained within the localized shifting
operators $\bsig(\rv)$ and $\bsig(\rv')$ on the right hand side. The
mere linear occurrence of the localized shifting operator (on the
right hand side) is remarkable and could arguably not have been
expected solely on the basis of the definition
\eqref{EQqsigDefinition}. To reveal the connection with Lie theory it
is useful to first consider the following modification.

The integrated shifting superoperator is given by multiplying the
localized version \eqref{EQqsigDefinition} with a given shifting field
$\eps(\rv)$ and integrating over position,
\begin{align}
  \Sigma[\eps] &= \int d\rv
  \eps(\rv)\cdot\bsig(\rv).
  \label{EQlongQuantumSigmaDefinition1}
\end{align}
The square brackets indicate functional dependence on the shifting
field $\eps(\rv)$, which is a $d$-dimensional smooth vector field.
The position integral~\eqref{EQlongQuantumSigmaDefinition1} can be
carried out explicitly via the commutator
definition~\eqref{EQqsigDefinition} and the definition of the one-body
current density operator \eqref{EQlongQuantumCurrentOperator}. The
result is
\begin{align}
  \Sigma[\eps]  &= -\frac{\rmi}{2\hbar}
  \sum_i[\,\cdot\,,  
    \eps(\rv_i)\cdot\hat\pv_i + \hat\pv_i\cdot\eps(\rv_i)],
  \label{EQlongQuantumSigmaDefinition2}
\end{align}
which depends clearly on the explicit form of $\eps(\rv)$.

Several properties of $\Sigma[\eps]$ are directly inherited from those
of the localized superoperator \eqref{EQqsigDefinition}. This includes
the application to the (Hilbert space) identity operator,
$\Sigma[\eps] 1 = 0$, and the (superoperator) anti-self-adjointness
$\Sigma^\dagger[\eps] = -\Sigma[\eps]$, as follows respectively from
the properties \eqref{EQqsigIdentiy} and \eqref{EQqsigAntiSelfAdjoint}
of the localized form.

The localized superoperator commutator
relationship~\eqref{EQqsigCommutator} is derived from the following
integrated version,
\begin{align}
  [\Sigma[\eps_1], \Sigma[\eps_2]] &= \Sigma[\eps_\Delta],
  \label{EQSigmaLieAlgebra}
\end{align}
where $\eps_1(\rv)$ and $\eps_2(\rv)$ are two, in general distinct
(smooth) shifting fields, see Appendix~\ref{APPderivationLie}.  The
`difference' shifting field $\eps_\Delta(\rv)$ that is the functional
argument on the right hand side of the integrated commutator
relationship~\eqref{EQSigmaLieAlgebra} is given by the following (also
smooth) difference:
\begin{align}
  \eps_\Delta(\rv) &=
  \eps_1(\rv)\cdot\nabla\eps_2(\rv) 
  - \eps_2(\rv)\cdot\nabla\eps_1(\rv),
  \label{EQepsDelta}
\end{align}
{and we recall the illustration shown in Fig.~\ref{FIG1}.}
Remarkably the right hand side of \eqr{EQepsDelta} constitutes the Lie
bracket of the two vector fields $\eps_1(\rv)$ and $\eps_2(\rv)$. Note
the presence of the minus sign on the above right hand side and that
the expression {\it does not} follow from any (naive) form of product
rule. Rather the Lie bracket of two vector fields is a staple of Lie
theory and a detailed account of the (elementary) geometrical
interpretation has been presented recently~\cite{mueller2024whygauge}.
While the presentation in Ref.~\cite{mueller2024whygauge} is framed in
a classical statistical mechanical context, we emphasize that the Lie
bracket transcends the classical and quantum many-body physics (see
Appendix~\ref{APPderivationLie}), as it is merely geometric in nature.

The role of the shifting field $\eps(\rv)$ is that of a mere gauge
function. As demonstrated \cite{mueller2025quantum} and summarized
below, the shifting leaves quantum statistical mechanical averages
invariant, such that the specific form of the (smooth) vector field
$\eps(\rv)$ is irrelevant.

\subsection{Equilibrium force and hyperforce sum rules}
\label{SECequilibriumSumRules}

We have so far described properties of the gauge transformation, based
on the localized shifting superoperator~\eqref{EQqsigDefinition}.  It
is now interesting to apply this concept to the standard framework of
quantum statistical mechanics for the description of many-body
Hamiltonians~\eqref{EQHamiltonian},
following Ref.~\cite{mueller2025quantum}. We start with operator
identities. Applying the localized shifting superoperator
\eqref{EQqsigDefinition} to the Hamiltonian~\eqref{EQHamiltonian}
yields
\begin{align}
  \hat\Fv(\rv) &=  -\bsig(\rv) \Hhat,
  \label{EQqsigHamiltonian} 
\end{align}
where the force density operator $\hat\Fv(\rv)$ is given by the
decomposition \eqref{EQlongQuantumForceDensityOperator} into kinetic,
interparticle potential and external potential contributions. Equation
\eqref{EQqsigHamiltonian} follows from explicit commutator
manipulations and it is compatible with the general physical concept
that forces arise from (negative) energy displacements.

In case of parametrically time-dependent Hamiltonians the
identity~\eqref{EQqsigHamiltonian} continues to hold in the
Schr\"odinger picture, which will be important for the dynamical gauge
invariance described below in
Sec.~\ref{SECdynamicalGaugeInvariance}. For the special case of
stationary Hamiltonians $\Hhat_0$ we have
\begin{align}
  \hat\Fv_0(\rv) &= -\bsig(\rv) \Hhat_0,
  \label{EQqsigInitialStateHamiltonian}
\end{align}
where $\hat \Fv_0(\rv)$ is the corresponding stationary force density
operator. When building the ensemble average and defining the thermal
mean $\Fv_0 = \langle \hat \Fv_0(\rv)\rangle$ it is straightforward to
obtain \cite{mueller2025quantum} from the identities
\eqref{EQqsigIdentiy} and \eqref{EQqsigAntiSelfAdjoint} the
equilibrium force density balance. This is exact and given by
\begin{align}
  \Fv_0(\rv) &= 0,
  \label{EQforceDensityBalance}
\end{align}
where the total force density on the left hand side consists of the
kinetic stress and the interparticle and external potential force
density contributions, as is inherited from the force density operator
sum \eqref{EQlongQuantumForceDensityOperator}.

We next apply the localized shifting superoperator
\eqref{EQqsigDefinition} to a given `hyperobservable' $\hat A$ and
obtain
\begin{align}
  \hat\Sv_A(\rv) = \bsig(\rv) \hat A,
  \label{EQqsigHyperforce}
\end{align}
which implies from the definition \eqref{EQqsigDefinition} that $\hat
\Sv_A(\rv) = -(\rmi/\hbar)[\hat A, m\hat\Jv(\rv)]$, which is identical
to the previous definition \eqref{EQlongQuantumSAOperatorPositions} of
the hyperforce density observable. Building the thermal mean of the
identity \eqref{EQqsigHyperforce} and using the anti-self-adjoint
superoperator property \eqref{EQqsigAntiSelfAdjoint} yields the
following quantum hyperforce sum rule~\cite{mueller2025quantum}:
\begin{align}
  \Sv_A(\rv) + \big( \hat A | \beta\hat\Fv_0(\rv) \big) &= 0,
  \label{EQhyperForceDensityBalance}
\end{align}
where the second term on the left hand side is the Mori product
\eqref{EQlongQuantumMoriProductDefinintion1} of the hyperobservable
$\hat A$ and the thermally scaled force density operator
\eqref{EQlongQuantumForceDensityOperator}. 

Due to the equilibrium force density balance
\eqref{EQforceDensityBalance} the occurring Mori product is identical
to the corresponding covariance, $\big( \hat A | \beta\hat\Fv_0(\rv)
\big)={\rm cov}\big( \hat A | \beta\hat\Fv_0(\rv) \big)$. Hence the
hyperforce density balance \eqref{EQhyperForceDensityBalance} can be
cast in the following alternative covariance form:
\begin{align}
  \Sv_A(\rv) + 
     {\rm cov}\big( \hat A | \beta\hat\Fv_0(\rv) \big) &= 0.
  \label{EQhyperForceDensityBalanceCovarianceForm}
\end{align}
For the present case of anti-self-adjoint operators $\hat A$ one can
reverse the order of arguments according to $\big( \hat A |
\beta\hat\Fv_0(\rv) \big) = \big(\beta\hat\Fv_0(\rv) | \hat A\big) \in
\mathbb{R}$ and similarly for the Mori covariance.  In this reversed
form the sum rules \eqref{EQhyperForceDensityBalance} and
\eqref{EQhyperForceDensityBalanceCovarianceForm} continue to hold for
general hyperobservables~$\hat A$, see Appendix~\ref{APPnonHermitian}.

Having summarized the main results of the equilibrium statistical
mechanical gauge invariance, we turn to results beyond those presented
in Ref.~\cite{mueller2025quantum}.

\subsection{Product and two-body  hyperforce sum rules}
\label{SEChigherSumRules}

The shifting superoperator \eqref{EQqsigDefinition} can be used in
flexible ways to derive exact sum rules. When applied to the product
$\hat A \hat B$ of two Hilbert space operators, one finds
\begin{align}
  \hat\Sv_{AB}(\rv)&=\bsig(\rv)\hat A\hat B\\&=
  \hat\Sv_A(\rv)\hat B + \hat A\hat\Sv_B(\rv),
\end{align}
which follows from the commutator form \eqref{EQqsigDefinition} of the
localized shifting operator $\bsig(\rv)$, the Leibniz (product) rule
for commutators $[\hat A \hat B, \hat C] = [\hat A, \hat C] \hat B +
\hat A [\hat B, \hat C]$ with the choice $\hat C= -(\rmi/\hbar) m \hat
\Jv(\rv)$, and expressing the hyperforce density operators
$\hat\Sv_A(\rv)$ and $\hat \Sv_B(\rv)$ via the commutator definition
\eqref{EQlongQuantumSAOperatorGeneral}.

From averaging over the statistical ensemble one obtains the following
`product rule' for hyperforce densities:
\begin{align}
  \Sv_{AB}(\rv) &=
  \langle \hat\Sv_A(\rv) \hat B \rangle +
  \langle \hat A \hat\Sv_B(\rv) \rangle,
  \label{EQlongQuantumProductHyperforceSumRule1}
\end{align}
where $\Sv_{AB} = \langle \hat \Sv_{AB}(\rv) \rangle$ is the product
hyperforce density distribution.  When applied to a product of two
operators the hyperforce sum rule \eqref{EQhyperForceDensityBalance}
becomes upon the trivial replacement of $\hat A$ by $\hat A\hat B$ the
following hyperforce product sum rule
\begin{align}
  \Sv_{AB}(\rv)
  + \big( \hat A \hat B | \beta \hat \Fv_0(\rv)   \big)
  &= 0.
  \label{EQlongQuantumProductHyperforceSumRule2}
\end{align}
The combination of the two above results
\eqref{EQlongQuantumProductHyperforceSumRule1} and
\eqref{EQlongQuantumProductHyperforceSumRule2} yields
\begin{align}
\langle
\hat\Sv_A(\rv)\hat B\rangle + \langle \hat A \hat\Sv_B(\rv) \rangle+
\big(\hat A\hat B|\beta\hat\Fv_0(\rv)\big)=0,
\end{align}
and we recall the one-body force density
operator~\eqref{EQlongQuantumForceDensityOperator} being that of the
equilibrium system, as is indicated by the subscript 0. The Mori
product can again be replaced equivalently by the Mori covariance
\eqref{EQlongQuantumMoriCovariance} due to general validity of the
equilibrium force density balance \eqref{EQforceDensityBalance}.

Higher-body correlation functions follow via averaging superoperator
products, such as $\bsig(\rv')\bsig(\rv)1=0$, as follows trivially
from the application \eqref{EQqsigIdentiy} to the identity
operator. The result is the following quantum analog of the classical
`3g'-sum rule \cite{sammueller2023whatIsLiquid}:
\begin{align}
  \big(\beta \hat \Fv_0(\rv)|\beta\hat \Fv_0(\rv')\big) + \langle \hat
      {\sf K}_0(\rv,\rv')\rangle=0,
\label{EQqsigThreeG}
\end{align}
where the Hamiltonian `curvature' operator is 
\begin{align}
  \hat {\sf K}_0(\rv,\rv') &=
  -\bsig(\rv)\bsig(\rv')\beta \Hhat_0,
\end{align}
which is equivalent to the force density gradient operator $\bsig(\rv)
\beta \hat\Fv_0(\rv')$ via \eqr{EQqsigInitialStateHamiltonian}.
Alternatively, \eqr{EQqsigThreeG} is obtained from the hyperforce sum
rule~\eqref{EQhyperForceDensityBalance} upon simply setting $\hat
A=\beta\hat\Fv_0(\rv')$ therein. Further multi-body sum rules follow
from the superoperator commutator structure~\eqref{EQqsigCommutator}.
An example is
\begin{align}
  & \big( \hat\Sv_A(\rv)| \beta \hat\Fv_0(\rv') \big)
  +\big( \hat\Sv_A(\rv') | \beta \hat \Fv_0(\rv) \big) \notag\\
  & \quad =
  [\nabla \delta(\rv-\rv')] \Sv_A(\rv)
  + \Sv_A(\rv') [\nabla \delta(\rv-\rv')].
  \label{EQcommutatorSumRule}
\end{align}
We next turn to the description of several interconnections and
generalizations of the operator shifting.

\section{Aspects of operator shifting}
\label{SECaspects}

We first explore the connection of operator shifting with quantum
canonical transformations
(Sec.~\ref{SECquantumCanonicalTransformation}) and then demonstrate
that operator shifting is compatible with quantum particle exchange
(Sec.~\ref{SECexchangeSymmetryAndShifting}). We then formulate
species-resolved shifting that is applicable to multi-component
systems (Sec.~\ref{SECmulticomponent}).

\subsection{Quantum canonical transformation}
\label{SECquantumCanonicalTransformation}

Quantum canonical transformations, as put forward by Anderson
\cite{anderson1994}, generalize their classical
counterparts~\cite{goldstein2002} and they correspond to unitary
transformations of Hilbert space \cite{anderson1994}.  Hermann and
Schmidt~\cite{hermann2022quantum} formulated the following quantum
canonical transformation of the fundamental position and momentum
operators:
\begin{align}
  \rv_i &\to \rv_i + \eps(\rv_i),
  \label{EQlongQuantumCanonicalTransformation1}\\
  \hat\pv_i &\to \frac{1}{2}\big\{
  \big[ \unity + \nabla_i \eps(\rv_i) \big ]^{-1} \cdot \hat\pv_i
  + \hat\pv_i \cdot \big[ \unity + \nabla_i \eps(\rv_i)\big]^{-\sf T} \big\}
  \label{EQlongQuantumCanonicalTransformation2}\\
  &\quad= \hat\pv_i -
  \frac{1}{2}\big\{ \big[\nabla_i \eps(\rv_i) \big] \cdot \hat\pv_i
  + \hat\pv_i \cdot \big[\nabla_i \eps(\rv_i)\big]^{\sf T} \big\}
  + \ldots
  \label{EQlongQuantumCanonicalTransformation3}
\end{align}
where we recall that the superscript $\sf T$ denotes transposing a
$d\times d$ matrix and $-\sf T$ indicates the combination of
transposition and matrix inversion. The form
\eqref{EQlongQuantumCanonicalTransformation3} arises from the Neumann
series in the gradient of the shifting field, applied to the general
momentum transform \eqref{EQlongQuantumCanonicalTransformation2}, and
writing out up to the linear order.

The interpretation of the position transform
\eqref{EQlongQuantumCanonicalTransformation1} is clearly that of
finite displacement. The momentum transform, both in the general form
\eqref{EQlongQuantumCanonicalTransformation2} and its linearized
version \eqref{EQlongQuantumCanonicalTransformation3}, is perhaps less
intuitive at first glance, but it serves the role of leaving the
canonical commutator relationships invariant under the joint
transform, see Ref.~\cite{hermann2022quantum}. From the invariance,
the validity of the equilibrium force density balance
\eqref{EQforceDensityBalance} follows
rigorously~\cite{hermann2022quantum}.

The present superoperator formalism performs an analogous role as the
quantum canonical transformation. We return to the integrated shifting
superoperator~\eqref{EQlongQuantumSigmaDefinition1} and apply
$1+\Sigma[\eps]$ to the position and momentum degrees of freedom of
one particle $i$. By explicit calculation one finds
\begin{align}
  (1+\Sigma[\eps])\rv_i &= \rv_i + \eps(\rv_i),
  \label{EQlongQuantumCanonicalTransformationSigma1}
  \\
  (1+\Sigma[\eps])\hat\pv_i &= \hat\pv_i
  -\frac{1}{2}\big\{
  \big[\nabla_i\eps(\rv_i)\big]\cdot\hat\pv_i
  + \hat\pv_i \cdot \big[\nabla_i \eps(\rv_i)\big]^{\sf T}\big\},
  \label{EQlongQuantumCanonicalTransformationSigma2}
\end{align}
where each of the two right hand sides is respectively identical to
the right hand side of
Eqs.~\eqref{EQlongQuantumCanonicalTransformation1} and
\eqref{EQlongQuantumCanonicalTransformation3}.  Hence the (linearized)
quantum canonical shifting transformation of
Ref.~\cite{hermann2022quantum} has an effect that is identical to the
present superoperator application, see
Appendix~\ref{APPquantumCanonical} for the demonstration of the
equivalence for a broad class of operators.

The significant advantage that lies in the formalization of shifting
via superoperators is that these allow one to identify clearly their
algebraic properties, as encapsulated in the commutator structure
\eqref{EQqsigCommutator} and \eqref{EQSigmaLieAlgebra}, and to
furthermore reveal the geometric interpretation in terms of the Lie
bracket \eqref{EQepsDelta}. The concrete connections with
  elementary geometry are described in the context of the classical
  setting in Ref.~\cite{mueller2024whygauge}; see in particular Sec.~3
  and the illustrative Fig.~5 therein.

\subsection{Exchange symmetry and shifting}
\label{SECexchangeSymmetryAndShifting}

Within our present formulation the exchange symmetry of the quantum
system is reflected in the properties of the Hilbert space basis
$|n\rangle$, which in position representation are wave functions
$\Psi_n(\rv_1,\ldots, \rv_N)$.  The exchange of two particles $i$ and
$j$ is then induced by the application of the transposition operator
${\cal P}_{ij}$, which is defined by the effect ${\cal P}_{ij}
\Psi(\rv_1,\ldots,\rv_N)= \Psi(\rv_1,\ldots,\rv_{i-1,}\rv_j,\rv_{i+1},
\ldots,\rv_{j-1},\rv_i,\rv_{j+1},\ldots \rv_N)$. The exchange symmetry
is then encapsulated in the induced change in sign in response to the
interchange of the two particles considered. In particular the
wavefunctions $\Psi(\rv_1,\ldots, \rv_N)$ satisfy anti-symmetry for
fermions, ${\cal P}_{ij}\Psi(\rv_1,\ldots,\rv_N) = -
\Psi(\rv_1,\ldots,\rv_N)$, and symmetry for bosons, ${\cal
  P}_{ij}\Psi(\rv_1,\ldots,\rv_N) = \Psi(\rv_1,\ldots,\rv_N)$, for all
indices $i$ and $j$.  Such basis functions can be constructed from
single-particle wave functions in the standard way, as described for
completeness in Appendix~\ref{APPoccupationNumber}.

The transposition operator ${\cal P}_{ij}$ can also be interpreted as
the superoperator which maps $\hat A$ to ${\cal P}_{ij} \hat A$, so
that ${\cal P}_{ij}$ denotes both the Hilbert space operator and the
superoperator. The application of the permutation superoperator has
the effect that $\rv_i \leftrightarrow \rv_j$ and $\hat\pv_i
\leftrightarrow \hat\pv_j$, so that particle indices of the
fundamental degrees of freedom are swapped, $i \leftrightarrow j$, see
Appendix~\ref{APPexchange} for more details.

Given that the shifting superoperator contains the summation of all
particles in its definition \eqref{EQqsigDefinition}, it is
straightforward to show that ${\cal P}_{ij} \bsig(\rv) = \bsig(\rv)
{\cal P}_{ij}$ and also ${\cal P}_{ij} \Sigma[\eps] = \Sigma[\eps]
{\cal P}_{ij}$, as the position integral
\eqref{EQlongQuantumSigmaDefinition1} has no effect on the exchange.
Hence the superoperator commutes with the localized shifting
superoperator,
\begin{align}
  [{\cal P}_{ij}, \bsig(\rv)] &= 0.
\end{align}
Similarly, for the integrated version, $[{\cal P}_{ij}, \Sigma[\eps]]
= 0$.  Hence operator shifting respects the exchange symmetry of the
quantum many-body system that is under considerations, see
Appendix~\ref{APPexchange} for the demonstration of compatibility with
general particle index permutations.

\subsection{Multi-component shifting}
\label{SECmulticomponent}
We provide an overview of the generalization of shifting to
multi-component systems with several species, indexed by
$\alpha=1,\ldots, M$, where $M$ denotes the number of components. It
is then convenient to introduce index sets ${\cal N}_\alpha$ that
contain those particle indices $i$ that constitute species $\alpha$.
The corresponding generalization of the one-component Hamiltonian
\eqref{EQHamiltonian} is standard,
\begin{align}
  \Hhat &= 
  \sum_\alpha \sum_{i \in {\cal N}_\alpha} \frac{\hat\pv_i^2}{m_\alpha}
  + u(\rv^N) 
  + \sum_\alpha \sum_{i \in {\cal N}_\alpha} V_{\rmext,\alpha}(\rv_i),
\end{align}
where the sums over $\alpha$ contain all $M$ species, $m_\alpha$
denotes the particle mass of species $\alpha$, and
$V_{\rmext,\alpha}(\rv)$ is the external potential acting on particles
of species $\alpha$. The interparticle potential $u(\rv^N)$ is
invariant under the exchange of two particles of the same species,
${\cal P}_{ij} u(\rv_1,\ldots,\rv_N)=u(\rv_1, \ldots, \rv_N)$ for all
indices $i,j\in {\cal N}_\alpha$.

The species-resolved one-body density and current density operators
are given, in generalization of the definitions
\eqref{EQlongQuantumDensityOperator} and
\eqref{EQlongQuantumCurrentOperator}, respectively by:
\begin{align}
  \hat\rho_\alpha(\rv) &= \sum_{i\in{\cal N}_\alpha} \delta(\rv-\rv_i),\\
  m\hat\Jv_\alpha(\rv) &= \frac{1}{2}\sum_{i\in {\cal N}_\alpha}
  [\hat\pv_i\delta(\rv-\rv_i)+\delta(\rv-\rv_i)\hat\pv_i].
\end{align}
Similarly, the species-resolved force density operator is
\begin{align}
  \hat\Fv_\alpha(\rv) &= \nabla\cdot\hat\taub_\alpha(\rv) 
  + \hat\Fv_{\rmint,\alpha}(\rv) 
  -\hat\rho_\alpha(\rv) \nabla V_{\rmext,\alpha}(\rv),
 \label{EQlongQuantumForceDensityOperatorMixture}
\end{align}
where the species-resolved kinetic stress density operator $\hat
\taub_\alpha(\rv)$ and the interparticle force density operator
$\hat\Fv_{\rmint,\alpha}(\rv)$ are respectively given by
Eqs.~\eqref{EQlongQuantumKineticStressOperator} and
\eqref{EQlongQuantumFintOperator} upon replacing $\hat\rho(\rv)$ by
$\hat\rho_\alpha(\rv)$ and restricting the further summations to
particle indices $i\in {\cal N}_\alpha$.

The species-resolved localized shifting superoperator follows as the
straightforward generalization of the one-component version
\eqref{EQqsigDefinition} as
\begin{align}
  \bsig_\alpha(\rv) &=
  -\frac{\rm i}{\hbar}[\,\cdot\,, m\hat\Jv_\alpha(\rv)].
\end{align}

In generalization of the commutator relationship
\eqref{EQqsigCommutator} for one-component systems one obtains
\begin{align}
  & [\bsig_\alpha(\rv),\bsig_\alphas(\rv')] =
  \notag\\&\qquad
  \delta_\aas[\nabla \delta(\rv-\rv')]\bsig_\alpha(\rv)
  +\delta_\aas\bsig_\alpha(\rv')[\nabla \delta(\rv-\rv')],
  \label{EQqsigCommutatorMixture}
\end{align}
where we recall $\delta_\aas$ as the Kronecker symbol.

The species-resolved hyperforce density operator is
\begin{align}
  \hat\Sv_{A,\alpha}(\rv) &= \bsig_\alpha(\rv) \hat A,
\end{align}
and the corresponding one-body hyperforce density is
$\Sv_{A,\alpha}(\rv) = \langle \hat\Sv_{A,\alpha}(\rv) \rangle$, which
generalize Eqs.~\eqref{EQqsigHyperforce} and
\eqref{EQlongQuantumSVAprofile}, respectively.  The one-body force
density distribution is defined as the thermal average
$\Fv_{\alpha}(\rv) = \langle \hat\Fv_{\alpha}(\rv) \rangle$. Then the
species-resolved force density balance is
\begin{align}
  \Fv_{\alpha}(\rv) &= 0,
\end{align}
and the species-resolved hyperforce sum rule reads as
\begin{align}
  \Sv_{A,\alpha}(\rv)
  + \big( \hat A | \beta \hat \Fv_\alpha(\rv) ) &= 0,
  \label{EQhyperForceDensityBalanceMixtures}
\end{align}
which constitute the respective generalization of the one-component
force density balance \eqref{EQforceDensityBalance} and hyperforce
density sum rule \eqref{EQhyperForceDensityBalance}.

\section{Hyperdensity functional theory}
\label{SEChyperDensityFunctionalTheory}

We first demonstrate consistency of the above gauge theory by showing
that the equilibrium hyperforce sum
rule~\eqref{EQhyperForceDensityBalance} can alternatively be derived
from the force density balance in a suitably extended ensemble
\cite{fick1990book, anero2013} (Sec.~\ref{SECextendedEnsemble}). The
extended ensemble then serves as the basis for construction of the
quantum hyperdensity functional theory, where we follow the lines of
argumentation of the classical counterpart
\cite{sammueller2024hyperDFT, sammueller2024whyhyperDFT} and lay out
the one-body minimization principle
(Sec.~\ref{SECquantumLongMinimizationPrinciple}), the
hyper-Ornstein-Zernike relation (Sec.~\ref{SECquantumLongHyperOZ}),
and general hyperobservables as density functionals
(Sec.~\ref{SECquantumLongHyperdensityFunctionals}).

\subsection{Extended ensemble and force balance}
\label{SECextendedEnsemble}

We consider the quantum statistical mechanics for the modified
Hamiltonian
\begin{align}
  \Hhat_{A}=\Hhat_0 -\lambda \hat A/\beta,
  \label{EQextendedHamiltonian}
\end{align}
where $\lambda$ is a coupling parameter.  We restrict ourselves to
(self-adjoint) forms of $\hat A$ that render $\Hhat_A$ well-behaved.  The
correspondingly extended quantum force density operator is generated
via the generic mechanism~\eqref{EQqsigHamiltonian} as
\begin{align}
  \hat \Fv_A(\rv) &= -\bsig(\rv)  \Hhat_A\\
  &= -\bsig(\rv) \Hhat_0 + \bsig(\rv) \lambda \hat A/\beta\\
  &=\hat\Fv_0(\rv) + \lambda \hat \Sv_A(\rv)/\beta,
\end{align}
where we have first made the extended
Hamiltonian~\eqref{EQextendedHamiltonian} explicit and then identified
both $\hat \Fv_0(\rv)$ and $\hat\Sv_A(\rv)$ via
Eqs.~\eqref{EQqsigInitialStateHamiltonian} and
\eqref{EQqsigHyperforce}, respectively. As the extended system is in
equilibrium, its mean force density vanishes,
\begin{align}
  \langle \hat\Fv_A(\rv) \rangle_A=0.
  \label{EQlongQuantumExtendedForceDensityBalance}
\end{align}
Here $\langle \, \cdot \, \rangle_A = \Tr \,\cdot\, \e^{-\beta
  \Hhat_A}/Z_A$ denotes the thermal average in the extended ensemble with
extended partition sum $Z_A = \Tr \e^{-\beta \Hhat_A}$. The Taylor
expansion of \eqr{EQlongQuantumExtendedForceDensityBalance} to linear
order in $\lambda$ yields
\begin{align}
  \langle \hat\Fv_A(\rv) \rangle_A
  &= \langle \hat\Fv_0(\rv)\rangle 
  + \frac{\lambda}{\beta}
  [\langle  \hat\Sv_A(\rv) \rangle 
    + {\rm cov}(\beta\hat\Fv_0(\rv)|\hat  A)]=0,
  \label{EQlongQuantumExtendedForceBalance1}
\end{align}
where all averages that occur in the Taylor expansion are now built in
the unmodified ensemble as is recovered for $\lambda=0$; see
Appendix~\ref{APPextendendEnsemble}. Clearly the coefficients of the
zeroth and first orders in $\lambda$ necessarily vanish separately,
\begin{align}
  & \langle \hat\Fv_0(\rv)\rangle  = 0,
  \label{EQlongQuantumExtendedForceBalance2}\\
  & \langle  \hat\Sv_A(\rv) \rangle 
    + {\rm cov}(\beta\hat\Fv_0(\rv)|\hat  A) = 0.
  \label{EQlongQuantumExtendedForceBalance3}
\end{align}
The zeroth order \eqref{EQlongQuantumExtendedForceBalance2}
constitutes the original force density balance
\eqref{EQforceDensityBalance} and the linear order
\eqref{EQlongQuantumExtendedForceBalance3} is the hyperforce sum
rule~\eqref{EQhyperForceDensityBalance}.

The present treatment continues to apply when going from the canonical
to the grand ensemble.  One merely needs to replace the canonical
trace $\Tr\,\cdot\,$ with the grand canonical analog, $\Tr'\,\cdot\,=
\sum_{N=0}^\infty\sum_n \langle n|\,\cdot\,|n\rangle e^{\beta\mu
  N}/N!$, where $\mu$ denotes the chemical potential. We will use the
grand ensemble for the quantum hyperdensity functional theory to which
we turn to next.

\subsection{Density functional minimization principle}
\label{SECquantumLongMinimizationPrinciple}

We formulate the quantum analog of the classical hyperdensity
functional theory by Samm\"uller {\it et
  al.}~\cite{sammueller2024hyperDFT, sammueller2024whyhyperDFT}.  We
work in the grand ensemble with the primed trace $\Tr'\,\cdot\,$ and
chemical potential $\mu$; see the definition of $\Tr'\,\cdot\,$ at the
end of the previous subsection. We consider the extended Hamiltonian
$\Hhat_A = \Hhat_0 - \lambda \hat A/\beta$, cf.~\eqr{EQextendedHamiltonian}.
We first make the hyperforce density sum rule in its covariance form
\eqref{EQhyperForceDensityBalanceCovarianceForm} more explicit by
inserting the decomposition of the force density operator
$\hat\Fv_0(\rv)$ into kinetic, interparticle, and external
contributions, see
Eqs.~\eqref{EQlongQuantumForceDensityOperator}--\eqref{EQlongQuantumFintOperator}.
This yields, upon using the Mori covariance,
\begin{align}
  &  \Sv_A(\rv) 
  +  \nabla\cdot {\rm cov}\big( \hat A |  \beta \hat\taub_0(\rv)\big)
  +  {\rm cov}\big( \hat A |  \beta \hat\Fv_{\rmint,0}(\rv)\big)
  \notag\\&\qquad\;\;
  - \chi_A(\rv) \nabla \beta V_{\rmext,0}(\rv) = 0,
  \label{EQhyperForceSumRuleSplitForm}
\end{align}
where $\chi_A(\rv)$ is the hyperfluctuation profile
\eqref{EQhyperFluctuationProfile}, which we reproduce for convenience,
\begin{align}
  \chi_A(\rv) &= {\rm cov}\big(\hat A | \hat\rho(\rv) \big),
  \label{EQhyperFluctuationProfileDFT}
\end{align}
and this will provide the link to density functional theory.  In the
re-writing \eqref{EQhyperForceSumRuleSplitForm} we have used that
$\nabla \cdot {\rm cov}(\hat A| \beta \hat\taub(\rv)) = {\rm cov}(\hat
A| \nabla \cdot \beta \hat \taub(\rv))$. For completeness the Mori
force covariance splitting is given explicitly by:
\begin{align}
  {\rm cov}(\hat A, \beta \hat\Fv_0(\rv))
  &= \nabla\cdot {\rm cov}\big( \hat A |  \beta \hat\taub_0(\rv)\big)
  +  {\rm cov}\big( \hat A |  \beta \hat\Fv_{\rmint,0}(\rv)\big)
  \notag\\&\quad
  - \chi_A(\rv) \nabla \beta V_{\rmext,0}(\rv).
  \label{EQforceCovarianceSplitting}
\end{align}

For the extended system, the grand potential density functional is
\begin{align}
  \Omega[\rho] &=
  F[\rho] + \int d\rv \rho(\rv)[V_{\rmext,0}(\rv) - \mu],
  \label{EQlongQuantumOmegaFunctional}
\end{align}
where $F[\rho]$ is the intrinsic free energy functional, which
contains kinetic, Hartree, exchange, correlation, and entropic
contributions \cite{mermin1965, kohn1999nobel}, as are generated by
the extended interparticle potential
\begin{align}
  u_A(\rv^N) &= u_0(\rv^N) - \lambda \hat A/\beta.
\end{align}

Mermin's minimization principle \cite{mermin1965} ascertains that
$\Omega[\rho]$ is minimized by the equilibrium density profile
$\rho_0(\rv)$ and that the value of the density functional at the
minimum is the true grand potential, $\Omega[\rho_0] = \Omega_0 =
-\beta^{-1}\ln \Tr'\e^{-\beta (\Hhat_A-\mu N)}$. At the minimum the
functional derivative vanishes,
\begin{align}
  \frac{\delta\Omega[\rho]}{\delta\rho(\rv)}\Big|_{\rho=\rho_0}=0
  \qquad \rm (min).
  \label{EQlongQuantumOmegaMinimum}
\end{align}
Calculating the functional derivative
\eqref{EQlongQuantumOmegaMinimum} on the basis of the decomposition
\eqref{EQlongQuantumOmegaFunctional} of the grand potential density
functional yields the Euler-Lagrange equation
\begin{align}
  C_1(\rv;[\rho]) - \beta V_{\rmext,0}(\rv) + \beta\mu &= 0.
  \label{EQEulerLagrange}
\end{align}
We have dropped the subscript 0 of the equilibrium density profile in
\eqr{EQEulerLagrange} and have defined the full one-body direct
correlation functional
\begin{align}
  C_1(\rv;[\rho]) &= -\frac{\delta \beta F[\rho]}{\delta\rho(\rv)},
  \label{EQlongQuantumC1definition}
\end{align}
which in particular includes the kinetic contributions.

\subsection{Hyper-Ornstein-Zernike relation}
\label{SECquantumLongHyperOZ}

Differentiating the Euler-Lagrange equation \eqref{EQEulerLagrange}
with respect to $\lambda$ yields, upon using the functional chain
rule, the following hyper-Ornstein-Zernike equation
\begin{align}
  C_A(\rv;[\rho]) + \int d\rv' C_2(\rv,\rv';[\rho]) \chi_A(\rv') &= 0,
  \label{EQhyperOrnsteinZernikeRelation}
\end{align}
where the hyperdirect correlation functional is given by
\begin{align}
  C_A(\rv;[\rho]) &=
  \frac{\partial C_1(\rv;[\rho])}{\partial \lambda}\Big|_\rho,
  \label{EQlongQuantumCAdefinition}
\end{align}
and the full two-body direct correlation functional is defined as
\begin{align}
  C_2(\rv,\rv';[\rho]) &=
  \frac{\delta C_1(\rv)}{\delta \rho(\rv')}.
\end{align}
The hyperfluctuation profile \eqref{EQhyperFluctuationProfileDFT}
emerges in \eqr{EQhyperOrnsteinZernikeRelation} due to the parametric
derivative relationship with the equilibrium density profile
\begin{align}
  \chi_A(\rv) = \frac{\partial  \rho(\rv)}{\partial \lambda},
\end{align}
as can be verified on the basis of the explicit average $\rho(\rv) =
\langle \hat\rho(\rv) \rangle_A$, taken in the extended grand
ensemble; we recall its setup described in
Sec.~\ref{SECextendedEnsemble} and refer to
Appendix~\ref{APPextendendEnsemble} for the explicit derivations.

\subsection{Hyperobservables as density functionals}
\label{SECquantumLongHyperdensityFunctionals}

Furthermore, parametric differentiation of the scaled grand potential
yields the mean of the considered observable~$\hat A$,
\begin{align}
  A &= -\frac{\partial \beta\Omega_0}{\partial \lambda},
\end{align}
as is described in Appendix~\ref{APPextendendEnsemble}.

The relevant functional relationships are as follows:
\begin{align}
  C_A(\rv;[\rho]) &= \frac{\delta A[\rho]}{\delta\rho(\rv)},
  \label{EQCAasFunctionalDerivative}\\
  A[\rho] &= \int {\cal D}[\rho] C_A(\rv;[\rho]),
  \label{EQmeanAasFunctionalIntegral}
\end{align}
where the functional line integral \cite{evans1992} in
\eqr{EQmeanAasFunctionalIntegral} is the inverse of
\eqr{EQCAasFunctionalDerivative}; the derivation of the latter
identity is shown below. Taking the limit $\lambda\to 0$ restores the
original ensemble with Hamiltonian~$\Hhat_0$. The present
argumentation is in formal analogy to the classical
case~\cite{sammueller2024hyperDFT, sammueller2024whyhyperDFT,
  sammueller2024multihyperDFT}.

It remains to derive the functional differentiation relationship
\eqref{EQCAasFunctionalDerivative}.  We start by recognizing that
\begin{align}
  A[\rho] =
  -\frac{\partial \beta \Omega[\rho]}{\partial \lambda}
  &= -\frac{\partial \beta   F[\rho]}{\partial \lambda}\Big|_\rho.
  \label{EQlongQuantumDFTderivation1}
\end{align}
The latter identity follows from the chain rule,
\begin{align}
  \frac{\partial \beta\Omega[\rho]}{\partial\lambda} &=
  \frac{\partial \beta F[\rho]}{\partial \lambda}\Big|_\rho
  + \int d\rv \chi_A(\rv)
  \frac{\delta\beta\Omega[\rho]}{\delta\rho(\rv)}\Big|_\lambda,
\end{align}
together with noting that the second term vanishes due to the density
functional minimization condition, $\delta
\Omega[\rho]/\delta\rho(\rv)|_\lambda=0$, as given by
\eqr{EQlongQuantumOmegaMinimum}.  We next differentiate both sides of
\eqr{EQlongQuantumDFTderivation1} with respect to the density profile.
This yields
\begin{align}
  \frac{\delta A[\rho]}{\delta \rho(\rv)}
  &=
  -\frac{\partial}{\partial \lambda}\Big|_\rho
  \frac{\delta \beta F[\rho]}{\delta\rho(\rv)}
  \label{EQlongQuantumDFTderivation4}\\
  &= \frac{\partial}{\partial \lambda}\Big|_\rho
  C_1(\rv;[\rho])
 = C_A(\rv;[\rho]),
\end{align}
where we have exchanged the order of differentiation, identified the
total one-body direct correlation functional $C_1(\rv;[\rho])$ by
\eqr{EQlongQuantumC1definition}, and used the definition of the
hyperdirect correlation functional $C_A(\rv;[\rho])$ via
\eqr{EQlongQuantumCAdefinition}.

\section{Dynamical gauge invariance}
\label{SECdynamicalGaugeInvariance}

We introduce temporal dependence in the quantum gauge theory along the
lines of argumentation of the classical dynamical gauge formulation
\cite{mueller2024dynamic}. The quantum theory is applicable to general
dynamics with parametrically time-dependent Hamiltonians. We start by
describing the details of this quantum dynamical setup
(Sec.~\ref{SECquantumLongManyBodyDynamics}) and then lay out the
dynamical operator shifting and the resulting nonequilibrium sum rules
(Sec.~\ref{SECquantumLongDynamicalOperatorShifting}).

\subsection{Microscopic many-body dynamics}
\label{SECquantumLongManyBodyDynamics}

We take the thermal equilibrium physics, generated by $\Hhat_0$, as
the initial state at time $t=0$ and consider the dynamics for $t\geq
0$ as induced by an, in general, explicitly time-dependent Hamiltonian
\eqref{EQHamiltonian}. The mass~$m$, the interparticle potential
$u(\rv^N)$, and the external potential $V_\rmext(\rv)$ can all depend
on time and we suppress such mere parametric time dependence in the
notation.  The quantum propagator $\calU(t,0)$ performs the time
evolution, which is unitary such that $\calU^\dagger(t,0)\calU(t,0) =
\calU(t,0)\calU^\dagger(t,0)=1$ and $\calU(0,0)=1$.  Heisenberg
operators are then given in the standard way as $\hat A(t) =
\calU^\dagger(t,0)\hat A\, \calU(t,0)$. The nonequilibrium statistical
mechanics is described by time-dependent averages that are built over
the ensemble of initial states, $A(t) = \langle \hat A(t) \rangle$.

\subsection{Dynamical operator shifting}
\label{SECquantumLongDynamicalOperatorShifting}

Generalizing \eqr{EQqsigDefinition} allows one to define the following
{\it dynamical} shifting superoperator:
\begin{align}
  \bsig(\rv,t) &= 
  -\frac{\rmi}{\hbar} [\,\cdot\,,m\hat\Jv(\rv,t) ]
  \label{EQqsigTime}
\end{align}
where the (scaled) Heisenberg one-body current density operator is
$m\hat\Jv(\rv,t)= \calU^\dagger(t,0) m\hat\Jv(\rv)\calU(t,0)$, see the
definition \eqref{EQlongQuantumCurrentOperator} of the corresponding
Schr\"odinger operator $m\hat\Jv(\rv)$. Several basic properties of
$\bsig(\rv,t)$ follow analogously to those of the static counterpart
$\bsig(\rv)$ described in Ref.~\cite{mueller2025quantum}.
Specifically one obtains in generalization of
Eqs.~\eqref{EQqsigIdentiy} and \eqref{EQqsigHamiltonian}: the trivial
identity $\bsig(\rv,t)1=0$ and the force density operator shifting
relationship $\hat\Fv(\rv,t) = -\bsig(\rv,t) \Hhat(t)$.  Furthermore
the dynamical hyperforce density operator is given by
\begin{align}
  \hat \Sv_A(\rv,t) &= \bsig(\rv,t) \hat A(t),
  \label{EQlongQuantumCurrentOperatorSVAdynamical}
\end{align}
where spelling out the right hand side yields the standard Heisenberg
form $\hat \Sv_A(\rv,t) = \calU^\dagger(t,0) \hat \Sv_A(\rv)
\calU(t,0)$ with $\hat\Sv_A(\rv)$ given by \eqr{EQqsigHyperforce}.
The dynamical shifting superoperator \eqref{EQqsigTime} retains much
of the favorable properties of its static counterpart: the
anti-self-adjointness~\eqref{EQqsigAntiSelfAdjoint} continues to hold,
\begin{align}
  \bsig^\dagger(\rv,t) &= -\bsig(\rv,t),
  \label{EQqsigAntiSelfAdjointDynamical}
\end{align}
and the dynamical trace identity is $\Tr \hat A [\bsig(\rv,t)\hat B] =
-\Tr [\bsig(\rv,t)\hat A] \hat B$, as follows from the general
derivation given in Appendix~\ref{APPinvarianceTrace}.

The implications of the dynamical gauge invariance reach beyond the
above generic Heisenberg time dependence. To reveal this structure, in
generalization of the initial state force density operator
\eqref{EQqsigInitialStateHamiltonian}, one applies $\bsig(\rv,t)$ to
the {\it initial state} Hamiltonian $\Hhat_0$.  Hence we define the
quantum `shift current operator' as
\begin{align}
  \hat\Cv(\rv,t) &= -\bsig(\rv,t) \beta \Hhat_0,
  \label{EQqsigShiftCurrent}
\end{align}
which is identical to the commutator form 
\begin{align}
  \hat\Cv(\rv,t) &=
  \frac{\rmi}{\hbar}[\beta \Hhat_0, m\hat\Jv(\rv,t)],
\end{align}
as is obtained from application of the explicit form of the dynamical
shifting superoperator \eqref{EQqsigTime} to $-\beta \Hhat_0$, which
we recall is the (scaled) Hamiltonian of the initial state. The shift
current operator~\eqref{EQqsigShiftCurrent} is a quantum observable,
$\hat \Cv(\rv,t)=\hat \Cv^\dagger(\rv,t)$, as is inherited from the
self-adjointness of the scaled current density operator,
$m\hat\Jv(\rv,t)=m\hat\Jv^\dagger(\rv,t)$, and preserved by the
definition~\eqref{EQqsigTime}, see Appendix~\ref{APPinvarianceTrace}.
The definition~\eqref{EQqsigShiftCurrent} is the quantum analog of the
classical hypercurrent observable, which is accessible in
trajectory-based simulations as an initial state time derivative
\cite{mueller2024gauge}.

The mean shift current, $\Cv(\rv,t)=\langle \hat \Cv(\rv,t) \rangle$,
satisfies the following exact shift current sum rule:
\begin{align}
  \Cv(\rv,t) &= 0,
  \label{EQqsigShiftCurrentIdentity}
\end{align}
where the left hand side can be decomposed into a sum of kinetic,
interparticle, and external contributions, as follows from the
corresponding splitting of $\Hhat_0$ in
\eqr{EQqsigShiftCurrent}. Briefly, \eqr{EQqsigShiftCurrentIdentity}
follows from averaging $\bsig(\rv,t)1=0$, such that $0 = \langle
\bsig(\rv,t) \rangle = \Tr \bsig(\rv,t) \e^{-\beta \Hhat_0}/Z =
(1|\hat\Cv(\rv,t))$ and recognizing the resulting Mori product as
$\langle \hat \Cv(\rv,t) \rangle$; see Appendix~\ref{APPcommutatorExp}
for the commutator of an exponentiated operator.  At the initial time,
the shift current operator~\eqref{EQqsigShiftCurrent} reduces to the
(scaled) equilibrium force density
operator~\eqref{EQqsigInitialStateHamiltonian}, such that
$\hat\Cv(\rv,0)=\beta\hat\Fv_0(\rv)$, and hence the shift current
identity \eqref{EQqsigShiftCurrent} becomes the force density
balance~\eqref{EQforceDensityBalance}.

The behaviour of general dynamical observables~$\hat A(t)$ follows
from generalizations of the derivations for the thermal equilibrium
case \cite{mueller2025quantum}. The dynamical shifting operator being
anti-self-adjoint leads to $\langle [\bsig(\rv,t)\hat
  A(t)]^\dagger\rangle = -\langle \hat A^\dagger(t)
\bsig(\rv,t)\rangle = -(\hat A(t)| \hat\Cv(\rv,t))$.  Identifying the
left hand side as the dynamical hyperforce density
$\Sv_A(\rv,t)=\langle \hat \Sv_A(\rv,t)\rangle=\langle \hat
\Sv_A^\dagger(\rv,t)\rangle$ and re-arranging yields the following
hypercurrent sum rule:
\begin{align}
  \Sv_A(\rv,t) + 
  \big(\hat A(t) | \hat\Cv(\rv,t)\big) &= 0,
  \label{EQqsigHypercurrentSumRule}
\end{align}
where $\hat\Cv(\rv,t)$ is given via \eqr{EQqsigShiftCurrent}.  The
second term in \eqr{EQqsigHypercurrentSumRule} measures via the Mori
product the correlation between the dynamical observable $\hat A(t)$
and the shift current density $\hat\Cv(\rv,t)$. In general this
average will be nonzero, despite the mean hypercurrent $\Cv(\rv,t)$
vanishing at all times, cf.\ the sum
rule~\eqref{EQqsigShiftCurrentIdentity}. For this reason, we can
equivalently express the Mori product in
\eqr{EQqsigHypercurrentSumRule} in Mori covariance form, which yields
the following alternative (Mori covariance) form of the hypercurrent
sum rule:
\begin{align}
  \Sv_A(\rv,t) + 
  {\rm cov}\big(\hat A(t) | \hat\Cv(\rv,t)\big) &= 0,
  \label{EQqsigHypercurrentSumRuleCovarianceForm}
\end{align}
and we refer to Appendix~\ref{APPnonHermitian} for the corresponding
hypercurrent sum rule for general (non-Hermitian) forms of~$\hat A$.

As a specific illustrative example, upon choosing $\hat A=\sum_i
\rv_i$ the general hyperforce density operator
\eqref{EQlongQuantumCurrentOperatorSVAdynamical} becomes
$\hat\Sv_A(\rv,t)=\hat\rho(\rv,t)\unity$. Thus the general
hypercurrent sum rule \eqref{EQqsigHypercurrentSumRuleCovarianceForm}
reduces to the following exact sum rule for the dynamical density
profile:
\begin{align}
  \rho(\rv,t) \unity &= -{\rm cov}\Big(\sum_i\rv_i(t)\Big |\hat
  \Cv(\rv,t)\Big).
\end{align}
In Appendix~\ref{APPrelationshipKuboIdentity} we give a brief
discussion of the relationship of the hypercurrent sum rule with a
standard Kubo identity \cite{fick1990book}.

\section{Dirac correspondence}
\label{SECquantumLongDiracCorrespondence}

It is useful to lay out the canonical quantization according to
Dirac's correspondence principle \cite{dirac1958book}. While the
generality of its validity is discussed and limitations have been
formulated systematically~\cite{groenewold1946}, we here use it to
relate the quantum and classical gauge invariances to each other. We
first describe the classical version of the many-body problem
(Sec.~\ref{SECquantumLongClassicalManyBodyDescription}) and then
relate and contrast the quantum and classical shifting transformations
to each other
(Sec.~\ref{SECquantumLongClassicalShiftingAndQuantumShifting}).

\subsection{Classical many-body model}
\label{SECquantumLongClassicalManyBodyDescription}

We start from a {\it classical} $N$-body system. The role of the
commutator is then played by the (scaled) Poisson brackets and general
classical observables $\hat A_\rmcl(\rv^N,\pv^N)$ are functions on the
$N$-body phase space. We use shorthand notations for positions,
$\rv^N=\rv_1,\ldots,\rv_N$, and for momenta,
$\pv^N=\pv_1,\ldots,\pv_N$, with $\pv_i$ denoting the momentum of the
classical particle $i=1,\ldots, N$, which is no longer an operator.
As a fundamental structure
\cite{zwanzig2001} given  two classical observables $\hat A_\rmcl$
and $\hat B_\rmcl$ their Poisson bracket is defined by
\begin{align}
  \{\hat A_\rmcl, \hat B_\rmcl\} &= \sum_i \Big(
    \frac{\partial \hat A_\rmcl}{\partial \rv_i}
    \cdot \frac{\partial \hat B_\rmcl}{\partial \pv_i}
    -    \frac{\partial \hat A_\rmcl}{\partial \pv_i}
    \cdot \frac{\partial \hat B_\rmcl}{\partial \rv_i}  \Big).
\end{align}

The Liouville equation for the time evolution of a classical
observable is $\partial \hat A_\rmcl(t)/\partial t = \{\hat
A_\rmcl(t), \Hhat(t)\}$, where the Hamiltonian is now a phase space
function. The time argument $t$ indicates dynamical dependence
according to the Heisenberg picture, which here is classical
\cite{zwanzig2001}.  As the phase space variables are canonical, they
satisfy $\{\rv_i,\rv_j\}=\{\pv_i,\pv_j\}=0$ and
$\{\rv_i,\pv_j\}=\delta_{ij}\unity$, where we recall $\delta_{ij}$ is
the Kronecker symbol and $\unity$ denotes the $d\times d$ unit matrix.
Choosing the observable of interest as the classical one-body current,
$\hat A = \hat\Jv_{\rm cl}(\rv)$, and applying the Liouville equation
of motion yields the classical force density phase space function
$\hat\Fv_{\rm cl}(\rv,t) = \{m\hat\Jv_{\rm cl}(\rv,t), \Hhat(t)\}$
\cite{schmidt2022rmp}.  The prior quantum description given in
Sec.~\ref{SECquantumLongManyBodyModel} then follows identically by
replacing Poisson brackets with (scaled) commutators, i.e.,
\begin{align}
  \{\,\cdot\,,\,\cdot\,\} &\to -\frac{\rmi}{\hbar}[\,\cdot\,,\,\cdot\,]
  \label{EQPoissonBracket}
\end{align}
and identifying classical phase space variables with their
corresponding quantum operators, $\hat A_\rmcl \to \hat A$.

We summarize briefly the classical statistical mechanics. Classical
thermal averages in the grand ensemble are given by $\langle \,\cdot\,
\rangle_\rmcl = \Tr'_\rmcl \,\cdot\, \e^{-\beta(H_\rmcl - \mu
  N)}/\Xi_\rmcl$, where $H_\rmcl$ is the classical version of the
quantum Hamiltonian~\eqref{EQHamiltonian} and the classical grand
partition sum is $\Xi_\rmcl = \Tr'_\rmcl \e^{-\beta(H_\rmcl-\mu
  N)}$. The classical `trace' is the following series of phase space
integrals $\Tr'_\rmcl = \sum_{N=0}^\infty (N!h^{dN})^{-1}\int
d\rv_1\ldots d\rv_N \int d\pv_1\ldots d\pv_N$ with Planck constant $h$
and spatial dimensionality $d$. The shifting gauge theory sketched in
the following remains valid in the canonical ensemble
\cite{robitschko2024any, mueller2024gauge, mueller2024whygauge}.

\subsection{Classical and quantum shifting}
\label{SECquantumLongClassicalShiftingAndQuantumShifting}

The recent {\it classical} statistical mechanical gauge theory
\cite{mueller2024dynamic} is based on viewing the above Poisson
brackets $\{m\hat\Jv_\rmcl(\rv,t),H_\rmcl(t)\}$ as an operator
$\{m\hat\Jv_\rmcl(\rv,t),\,\cdot\,\}$ that acts on the classical
version of the Hamiltonian $H_\rmcl(t)$. We consider the static case
in the following. Upon exchanging the order of arguments and
multiplying by $-1$, one obtains the following classical `shifting'
operators \cite{mueller2024gauge, mueller2024whygauge,
  mueller2024dynamic}, which are given in classical Schr\"odinger form
\cite{zwanzig2001} as:
\begin{align}
  \bsig_\rmcl(\rv) &= \{\,\cdot\,,m\hat\Jv_\rmcl(\rv)\},
  \label{EQbsigClassicalPoisson}
\end{align}
and we refer to Ref.~\cite{mueller2024dynamic} for the explicit phase
space form.  The classical shifting operators
\eqref{EQbsigClassicalPoisson} represent a canonical transformation
\cite{goldstein2002} and they generate a specific gauge transformation
on phase space \cite{mueller2024gauge, mueller2024whygauge}.  The
shifting operators possess remarkable (algebraic) properties, such as
being anti-self-adjoint on phase space and having nontrivial (Lie
algebra) commutator structure.  This differential operator structure
has profound consequences when applied to classical statistical
mechanical averages both in thermal equilibrium
\cite{mueller2024gauge, mueller2024whygauge} and in general
nonequilibrium situations \cite{mueller2024dynamic}. A range of exact
sum rules follows and these have been shown to be computationally
accessible using classical particle-based simulations
\cite{robitschko2024any, mueller2024dynamic}.

Examples of the use of the classical shifting operators are their
application to the classical Hamiltonian $H_\rmcl$ and a given
classical hyperobservable $\hat A$, which we recall to be a phase
space function in the present classical context. The result is the
classical force density observable \cite{schmidt2022rmp,
  mueller2024gauge}, $\hat\Fv_{\rmcl}(\rv) = -\bsig_\rmcl(\rv)
H_\rmcl,$ and the classical hyperforce density phase space
function~\cite{robitschko2024any, mueller2024gauge}, $\hat \Sv_{A,
  \rmcl}(\rv) = \bsig_\rmcl(\rv) \hat A$.  These relationships are in
formal analogy to the application of the quantum superoperator
\eqref{EQqsigDefinition} to the (negative) Hamiltonian, which gives
the force density operator \eqref{EQqsigHamiltonian}, and to a general
hyperobservable, which yields the hyperforce density operator
\eqref{EQqsigHyperforce}. We recall that in more elementary form the
quantum hyperforce density operator $\hat\Sv_A(\rv)$ is given for
position-only dependent hyperobservables $\hat A(\rv^N)$ by
\eqr{EQlongQuantumSAOperatorPositions} and for more general
observables by \eqr{EQlongQuantumSAOperatorGeneral}.

We can now observe that the Dirac correspondence principle relates the
classical phase space shifting operator \eqref{EQbsigClassicalPoisson}
to the quantum shifting superoperator \eqref{EQqsigDefinition}.
Remarkably, the Dirac correspondence is passed down to the form of the
resulting sum rules, where a representative example is the classical
hyperforce density balance,
\begin{align}
  \Sv_A(\rv) + 
     {\rm cov}(\hat A_\rmcl, \beta \hat\Fv_\rmcl(\rv))
     & = 0,
\end{align}
where the covariance of two classical phase space functions is defined
as ${\rm cov(\hat A_\rmcl, \hat B_\rmcl)} = \langle \hat A_\rmcl \hat
B_\rmcl \rangle_\rmcl - A_\rmcl B_\rmcl$, with the classical averages
of the two observables being given by $A_\rmcl = \langle \hat A_\rmcl
\rangle_\rmcl$ and $B_\rmcl = \langle \hat B_\rmcl \rangle_\rmcl$.

The correspondence on the level of the gauge invariance sum rules
remains nontrivial though, as is manifest by the occurrence of the
Mori product (and covariance) in the quantum version, which any
`naive' replacement rule would {\it not} pick up. Moreover, we find it
highly remarkable that the intricacies of the correct implementation
of the superoperator calculus are compatible with the general gauge
shifting concept.

\section{Conclusions}
\label{SECquantumLongConclusions}

In conclusion, we have complemented the presentation of the quantum
statistical mechanical gauge theory of Ref.~\cite{mueller2025quantum}
by i)~providing further background on the particle-based quantum
mechanical underpinnings, ii)~exploring further consequences of the
shifting both in and out of equilibrium, and iii)~laying out
theoretical interconnections with an extended quantum density
functional approach, which is analogous to the classical hyperdensity
functional theory~\cite{sammueller2024hyperDFT,
  sammueller2024whyhyperDFT}, that applies to general observables.
The shifting gauge transformation possesses geometric character and it
is represented by a quantum `superoperator', i.e., a map between
standard Hilbert space operators. We have given background for several
mathematical concepts that are required to manipulate superoperators,
such as building the adjoint and forming the commutator of two
superoperators.

The specific form of the shifting superoperator, see its definition
via the commutator \eqref{EQqsigDefinition} with the scaled current
density operator, induces nontrivial Lie algebra structure that
characterizes the shifting. In particular the relationship between two
shifting superoperators via the commutator relationship
\eqref{EQSigmaLieAlgebra} is noteworthy due to the occurrence of the
Lie bracket \eqref{EQepsDelta} of the two shifting vector fields that
are involved. The corresponding spatially localized distributional
commutator~\eqref{EQqsigCommutator} forms a similarly efficient point
for the sum rule construction.
Besides the equilibrium hyperforce sum rule
\eqref{EQhyperForceDensityBalance} and the further results that are
presented in Ref.~\cite{mueller2025quantum}, we have here obtained
product and two-body hyperforce sum rules, see
Sec.~\ref{SEChigherSumRules}, as well as the species-resolved
hyperforce balance \eqref{EQhyperForceDensityBalanceMixtures} that
applies to multi-component systems. We have generalized the shifting
gauge invariance framework to nonequilibrium situations that are
induced by parametric time dependence of the Hamiltonian. One key
result of these considerations of the many-body dynamics is the exact
nonequilibrium hypercurrent sum
rule~\eqref{EQqsigHypercurrentSumRule}.

The equilibrium hyperforce framework connects naturally to
hyperdensity functional theory, which we have formulated to generalize
standard density functional dependence. Instead of targeting the grand
potential, or equivalently the groundstate energy in the
low-temperature limit, here the thermal mean~$\langle \hat A \rangle$
of any given hyperobservable $\hat A$ is expressed as an explicit
density functional $A[\rho]$, see
Sec.~\ref{SEChyperDensityFunctionalTheory}.
An explicit bridge between the quantum gauge theory and the
hyperdensity formulation is provided by the spatially resolved
hyperfluctuation profile $\chi_A(\rv)$. Notably $\chi_A(\rv)$ features
in both the split form of the hyperforce balance sum rule
\eqref{EQhyperForceSumRuleSplitForm} as well as in the
hyper-Ornstein-Zernike equation
\eqref{EQhyperOrnsteinZernikeRelation}.
Relevant background and detailed derivations of all our key results
are given in the Appendices, which we summarize below.

As an outlook on the potential for future work, it could be highly
interesting to carry out numerical investigations of the gauge
correlation structure for specific systems of interest. Clearly, this
task requires careful choices of suitable systems and also to have a
handle on a numerically exact or approximate representation of the
quantum many-body physics. Such a method must allow one both to
perform the involved Hilbert space operator calculus and to carry out
the thermal average, either to put the equilibrium gauge structure to
work, as described in Sec.~\ref{SECequilibriumGaugeInvariance}, or to
average over the initial state ensemble that is implied by the
dynamical setup, see Sec.~\ref{SECdynamicalGaugeInvariance}.
Some inspiration for specific choices of systems could come from
corresponding work carried out for classical systems, where a range of
concrete choices of hyperobservables was investigated on the basis of
particle-based simulation work \cite{robitschko2024any,
  sammueller2024hyperDFT, sammueller2024whyhyperDFT, matthes2024mix}
in spatially confined systems \cite{robitschko2024any, matthes2024mix}
and for cluster formation \cite{sammueller2024hyperDFT,
  sammueller2024whyhyperDFT, sammueller2024multihyperDFT}.

Whether the present framework can serve to develop machine learning
schemes that address the many-body problem \cite{carleo2017,
  huang2023review} is an interesting question. Quantum density
functional theory is a competitive contender for providing a sound
platform \cite{huang2023review} and our present framework supplies
both the existence of the required functional relationships, which
constitute unique maps from the density profile to the averaged
hyperobservables, as well as exact sum rules that could form the basis
for consistency checks and to regularize learning; we refer to
Refs.~\cite{sammueller2024pairmatching, kampa2026pairmatching} for
regularized classical density functional learning and to
Refs.~\cite{sammueller2023neural} for using classical sum rules as
consistency checks for neural density functionals.
Finding connections to modern electronic density functional approaches
\cite{tokatly2005one, tokatly2005two, tokatly2007, ullrich2006,
  tchenkoue2019, tarantino2021, ullrich2025} is worthwhile, including
time-dependent density functional theory \cite{ullrich2025, daas2025}
and machine learning methodology~\cite{nagai2023,akashi2025}.

The present gauge theory constitutes arguably a flexible toolbox for
both defining and investigating correlation functions that follow from
first principles, such as the hyperforce density distribution
$\Sv_A(\rv)$, the Mori force covariance ${\rm cov}(\hat A
|\beta\hat\Fv(\rv))$ with its three individual
constituents~\eqref{EQforceCovarianceSplitting}, the hyperfluctuation
profile~$\chi_A(\rv)$, and the hyperdirect
correlation~function~$c_A(\rv)$.

Whether the gauge shifting can shed light on the topical problem of
many-body localization in bosonic systems \cite{choi2016, bordia2017,
  yan2017prl,yan2017pra} is an interesting question.
We recall that our present setup is based on the assumption of an
initally thermalized system, which can undergo a sudden change via
explicit time dependence of a tuning parameter in the Hamiltonian,
such that the system is pushed away from equilibrium.  Whether then
thermalization occurs is a relevant question, see
e.g.~Refs.~\cite{lydzba2023, patil2026} for recent work and
Ref.~\cite{patil2026review} for an introduction.
Also addressing orientational degrees of freedom, as was performed
classically \cite{hermann2021noether, nguyen2026}, and investigating
the interplay with standard symmetries \cite{phamvan2026symmetry}, as
well as relating to a mathematically rigorous distributional
formulation \cite{maruyama2026} can be relevant tasks.

{The very recent work by Pham-Van \cite{phamvan2026quantumGeometry,
    phamvan2026exchangePrivateLindblad} on quantum statistical gauge
  invariance is noteworthy as it points strongly towards the relevance
  of the framework. The study \cite{phamvan2026quantumGeometry}
  presents a geometric organization of the quantum shift symmetry,
  together with investigations of an entire spectrum of deep
  relationships with a broad range of topics in condensed
  matter. Furthermore several key identities are verified using exact
  numerical diagonalization techniques.}

The seven
Appendices~\ref{APPinnerProducts}--\ref{APPrelationshipKuboIdentity}
are organized as follows.
Details and background for the Hilbert-Schmidt and Mori inner products
are given in Appendix~\ref{APPinnerProducts}.
The proof of the invariance of the trace under cyclic permutations and
the associated commutator superoperators is shown in
Appendix~\ref{APPinvarianceTrace}.
We give the explicit derivation of the Lie algebra commutator
relationships in Appendix~\ref{APPderivationLie}.
The commutator of two shifting superoperators is described in
Appendix~\ref{APPcommutatorShiftingOperators}.
We lay out the hyperforce and hypercurrent balance for non-Hermitian
operators in Appendix~\ref{APPnonHermitian}.
The relationship of the quantum canonical transformation with the
shifting superoperator is described in
Appendix~\ref{APPquantumCanonical}.
The description on the basis of occupation numbers is described in
Appendix~\ref{APPoccupationNumber}.
The compatibility of operator shifting with particle index
permutations is proven Appendix~\ref{APPexchange}.
Background for the description of thermal averages in the extended
thermal average is provided in Appendix~\ref{APPextendendEnsemble}.
The commutator with an exponentiated operator is described in
Appendix~\ref{APPcommutatorExp}.
The relationship of the hypercurrent sum rule with a standard Kubo
identity is discussed in Appendix~\ref{APPrelationshipKuboIdentity}.

\begin{acknowledgments}

We thank Florian Samm\"uller, Robert Evans, Kieron Burke, and Hai Pham
Van for useful discussions. Hai Pham Van is also gratefully
acknowledged for providing us with his independently achieved results
on exchange symmetry \cite{phamvan2026exchangePrivate}.  This work is
supported by the DFG (Deutsche Forschungsgemeinschaft) under Project
No.~551294732.

\end{acknowledgments}

\bibliographystyle{prsty}
\bibliography{noe}

	\appendix

	\section{The Hilbert-Schmidt inner product and the Mori product}
	\label{APPinnerProducts}
	
	We start by laying out some background. The $N$-body Hilbert space 
 is equipped with the standard inner product $\langle n \vert n^\prime \rangle = \int d \b r^N (n(\b r^N))^\star n^\prime(\b r^N)$. Important physical restrictions of the general Hilbert space concept include using only functions which are either symmetric or anti-symmetric under particle exchanges to describe the respective statistics of (indistinguishable) bosons and fermions, as discussed in Sec.~\ref{SECexchangeSymmetryAndShifting} of the main text.
	
	The space of linear Hilbert space operators, sometimes referred to as Liouville space~\cite{fick1990book}, forms a complex vector space which can be equipped with some inner product. A standard example is the Hilbert-Schmidt inner product~\cite{fick1990book} given by
	\begin{align} \label{eq_standard_inner_prod}
		\Tr \hat A^\dagger \hat B.
	\end{align}
	Note that the quantum trace $\Tr \cdot = \sum_{n} \langle n \vert \cdot \vert n \rangle$ of the Hilbert space is independent of the chosen orthonormal basis $\vert n \rangle$~\cite{fick1990book}.
	
	The mapping~\eqref{eq_standard_inner_prod} is linear in the second argument. Furthermore, it is positive-definite and conjugate symmetric, since $\Tr \hat A^\dagger \hat A = \sum_n \langle \hat A n \vert \hat A n\rangle > 0$ for $\hat A \neq 0$ and $(\Tr \hat A^\dagger \hat B)^{\star} = \sum_{n} \langle \hat A n \vert \hat B n \rangle^{\star} = \sum_{n} \langle \hat B n \vert \hat A n \rangle = \Tr \hat B^\dagger \hat A$, where we have used that the Hilbert space inner product $\langle \ccdot \vert \ccdot \rangle$ is positive definite and conjugate symmetric. Thus the Hilbert-Schmidt inner product~\eqref{eq_standard_inner_prod} is indeed an inner product~\cite{fick1990book}, which hence serves as a valid starting point in Sec.~\ref{SECequilibriumShifting} to define the adjointness of superoperators.
	
	The inner product~\eqref{eq_standard_inner_prod} of two linear operators is not unique and the Mori product $(\ccdot|\ccdot)$ defined in Eq.~\eqref{EQlongQuantumMoriProductDefinintion1} forms a further alternative version. Due to the invariance of the quantum trace under cyclic permutations, see Appendix~\ref{APPinvarianceTrace} below, the inner product~\eqref{eq_standard_inner_prod} and the Mori product~\eqref{EQlongQuantumMoriProductDefinintion1} are linked via $(\hat A | \hat B) = \beta^{-1} \int_0^\beta d\beta' \Tr \hat A_{\beta^\prime}^\dagger \hat B_{\beta^\prime}/Z$ where $\hat C_{\beta^\prime} = \e^{-\beta^\prime \Hhat_0/2} \hat C \e^{(\beta^\prime-\beta)\Hhat_0/2}$ for $\hat C = \hat A, \hat B$. Despite their connection, both inner products differ in general, i.e., each one yields a different complex number.
	
	The properties of the inner product~\eqref{eq_standard_inner_prod} imply that the Mori product is also linear in the second argument and positive-definite, since $\Tr \hat A_{\beta^\prime}^\dagger \hat A_{\beta^\prime} > 0$ with $\hat A_{\beta^\prime} \neq 0 $ and thus $(\hat A \vert \hat A) > 0$ for $\hat A \neq 0$. The conjugate symmetry of the Mori product follows from the conjugate symmetry of the inner product~\eqref{eq_standard_inner_prod}, since $(\Tr \hat A_{\beta^\prime}^\dagger \hat B_{\beta^\prime})^\star = \Tr \hat B_{\beta^\prime}^\dagger \hat A_{\beta^\prime}$ and thus~\cite{fick1990book}
	\begin{align} \label{eq_mori_conj_symm}
		(\hat A \vert \hat B)^\star = (\hat B \vert \hat A).
	\end{align}
	This proves that the Mori product indeed constitutes an inner product~\cite{fick1990book}.
	
	The Mori product has further useful properties which we summarize briefly. Considering adjoint operators yields~\cite{fick1990book}
	\begin{align}
		(\hat A \vert \hat B) = (\hat B^\dagger \vert \hat A^\dagger), \label{eq_mori_adj}
	\end{align}
	which follows by standard adjoint argumentation and the invariance of the quantum trance under cyclic permutation, see Appendix~\ref{APPinvarianceTrace}, since $(\hat B^\dagger_{\beta^\prime})^\dagger = \hat B_{\beta^\prime}$ and $\Tr \hat B_{\beta^\prime} \hat A^\dagger_{\beta^\prime} = \Tr \hat A^\dagger_{\beta^\prime} \hat B_{\beta^\prime}$. Combining Eqs.~\eqref{eq_mori_conj_symm} and~\eqref{eq_mori_adj} yields the relation~\cite{fick1990book} $(\hat A \vert \hat B) = (\hat A^\dagger \vert \hat B^\dagger)^{\star}$, which as a sideeffect proves that the Mori product of self-adjoint operators is indeed real, as discussed in Sec.~\ref{SECquantumLongManyBodyModel}.

	\section{The invariance of the trace under cyclic permutations and commutator superoperators}
	\label{APPinvarianceTrace}
	
	A particularly useful property of the quantum trace is its invariance under cyclic permutations~\cite{fick1990book, zwanzig2001}: Using twice the identity $1 = \sum_{m} \vert m \rangle \langle m \vert$ yields
	\begin{align}
		\Tr \hat A \hat B \hat C &= \sum_{n} \langle n \vert \hat A \hat B \hat C \vert n \rangle \\
		&= \sum_{n,m,k} \langle n \vert \hat A m \rangle \langle m \vert \hat B k \rangle \langle k \vert \hat C n \rangle. \label{eq_quant_trace_inv_deriv}
	\end{align}
	Since the sum~\eqref{eq_quant_trace_inv_deriv} is trivially invariant under cyclic permutations of the three factors, we conclude
	\begin{align} \label{eq_quant_trace_inv}
		\Tr \hat A \hat B \hat C = \Tr \hat B \hat C \! \hat A = \Tr \hat C \! \hat A \hat B.
	\end{align}
	
	The invariance~\eqref{eq_quant_trace_inv} of the quantum trace under cyclic permutations has certain implications. For example setting $\hat C = 1$ leads to $\Tr \hat A \hat B = \Tr \hat B \hat A$ and thus~\cite{fick1990book}
	\begin{align}
		\Tr [\hat A, \hat B] = 0. \label{eq_quant_trace_comm}
	\end{align}
	The commutator product rule implies that $\hat A [\hat B, \hat C] = [\hat A \hat B, \hat C] - [\hat A, \hat C] \hat B$, where the first term on the right-hand side vanishes when building the quantum trace, $\Tr [\hat A \hat B, \hat C] = 0$, due to Eq.~\eqref{eq_quant_trace_comm}. The anti-symmetry of the commutator then yields the following commutator trace identity~\cite{fick1990book}
	\begin{align} \label{eq_quant_trace_prod}
		\Tr \hat A [\hat B, \hat C] = \Tr [\hat C, \hat A] \hat B.
	\end{align}
	
	As a consequence of Eq.~\eqref{eq_quant_trace_prod}, superoperators $\Ocal_C$ of the form $\Ocal_C = [\ccdot, \hat C]$ satisfy
	\begin{align} \label{eq_superop_trace}
		\Tr \hat A (\Ocal_C \hat B) = - \Tr (\Ocal_C \hat A) \hat B.
	\end{align}
	Furthermore, the property of adjoint commutators $[\hat A^\dagger, \hat C] = - [\hat A, \hat C^\dagger]^\dagger$  implies 
	\begin{align} \label{eq_superop_adjoint}
		\Ocal_C \hat A^\dagger = - (\Ocal_C^\dagger \hat A)^\dagger,
	\end{align}
	where $\Ocal_C^\dagger$ is defined as $\Ocal_C^\dagger = [\ccdot, \hat C^\dagger]$. Replacing $\hat A$ with $\hat A^\dagger$ in Eq.~\eqref{eq_superop_trace} and using Eq.~\eqref{eq_superop_adjoint} leads to 
	\begin{align} \label{eq_superop_trace_adjoint}
		\Tr \hat A^\dagger (\Ocal_C \hat B) = \Tr (\Ocal_C^\dagger \hat A)^\dagger \hat B,	
	\end{align}
	so $\Ocal_C^\dagger = [\ccdot, \hat C^\dagger] $ is indeed  the adjoint superoperator of $\Ocal_C$.
	
	Note that the shifting superoperator~\eqref{EQqsigDefinition} can be written as $\bg \sigma(\b r) = \Ocal_C$ with $\hat C = -(\ii/\hbar) m \bhat J(\b r)$ and $\hat C^\dagger = - \hat C$, so the shifting superoperator satisfies Eqs.~\eqref{eq_superop_trace}--\eqref{eq_superop_trace_adjoint} as discussed in Ref.~\cite{mueller2025quantum}. Since the dynamical shifting superoperator~\eqref{EQqsigTime} has the form $\bg \sigma(\b r,t) = \Ocal_C$ with $\hat C = -(\ii/\hbar) m \bhat J(\b r,t)$, the argumentation can be transferred directly to the dynamical setting.

	\section{Derivation of the Lie algebra commutator relation}
	\label{APPderivationLie}
	In order to derive the exact form of the difference shifting field $\eps_\Delta(\b r)$, see Eq.~\eqref{EQepsDelta}, we define for the sake of brevity the integrated (scaled) current density $\hat \Jcal [\eps] = \sum_i [\bhat p_i \cdot \eps(\b r_i) + \eps(\b r_i) \cdot \bhat p_i]/2$. Hence the integrated shifting superoperator~\eqref{EQlongQuantumSigmaDefinition1} has the form $\Sigma[\eps] = -(\ii/\hbar)[\ccdot, \hat \Jcal[\eps]]$.
	
	The commutator $[\ccdot, \ccdot]$ statisfies the Jacobi identity, i.e. $[[\hat A, \hat B], \hat C] + [[\hat B, \hat C], \hat A] + [[\hat C, \hat A], \hat B] = 0$ for given Hilbert space operators $\hat A$, $\hat B$ and $\hat C$, so that $[[\ccdot, \hat A], [\ccdot, \hat B]] = [[\ccdot, \hat B], \hat A] - [[\ccdot, \hat A], \hat B] = [[\ccdot, \hat B], \hat A] + [[\hat A, \ccdot], \hat B] = - [[\hat B, \hat A], \ccdot] = -[\ccdot ,[\hat A, \hat B]]$, where we have exploited the anti-symmetry and the Jacobi identity of the commutator. When using the superoperator notation of Appendix~\ref{APPinvarianceTrace}, the above relation can be expressed equivalently as $[\Ocal_A, \Ocal_B] = \Ocal_C$ with $\hat C = - [\hat A, \hat B]$, so we conclude that
	\begin{align}
		[\Sigma[\eps_1], \Sigma[\eps_2]] &= \Big[\Big[\ccdot, \Big(-\frac{\ii}{\hbar} \hat \Jcal[\eps_1]\Big) \Big], \Big[\ccdot, \Big(-\frac{\ii}{\hbar} \hat \Jcal[\eps_2]\Big)\Big]\Big] \notag \\
		&= -\frac{\ii}{\hbar} \Big[\ccdot, \frac{\ii}{\hbar} \Big[\hat \Jcal[\eps_1], \hat \Jcal[\eps_2]\Big]\Big]. \label{eq_deriv_eps_delt_jacobi}
	\end{align}
	It thus remains to analyze $(\ii/\hbar) [\hat \Jcal[\eps_1], \hat \Jcal[\eps_2]]$ in more detail which we do in the following.
	
	Since the fundamental degrees of freedom of different particles commute, the double sum that occurs in the commutator simplifies to
	\begin{align}
		&(\ii/\hbar) [\hat \Jcal[\eps_1], \hat \Jcal[\eps_2]] \notag \\
		&= \frac{\ii}{4\hbar} \sum_{i,j} [\bhat p_i \cdot \eps_1(\b r_i) + \eps_1(\b r_i) \cdot \bhat p_i, \bhat p_j \cdot \eps_2(\b r_j) + \eps_2(\b r_j) \cdot \bhat p_j] \notag \\
		&= \frac{\ii}{4\hbar} \sum_{i} [\bhat p_i \cdot \eps_1(\b r_i) + \eps_1(\b r_i) \cdot \bhat p_i, \bhat p_i \cdot \eps_2(\b r_i) + \eps_2(\b r_i) \cdot \bhat p_i], \label{eq_deriv_eps_delt_one_part}
	\end{align}
	where the re-written form only involves a single sum.
	
	Thus we consider the argument of the sum with generic position $\b r$ and generic momentum $\bhat p = - \ii \hbar \nabla$. By using the `generalized' canonical commutation relations $[\bhat p, \bhat p] = 0$, $[\bhat p, \eps(\b r)] = -\ii \hbar [\nabla \eps(\b r)]$, and $[\eps(\b r), \tilde {\eps}(\b r)] = 0$, we simplify and conclude:
	\begin{align}
		&\frac{\ii}{4 \hbar}[\bhat p \cdot \eps_{1}(\b r) + \eps_{1}(\b r) \cdot \bhat p, \bhat p \cdot \eps_{2}(\b r) + \eps_{2}(\b r) \cdot \bhat p] \notag \\
		&=\frac{\ii}{2 \hbar} \big[\bhat p \cdot \big(\eps_{1}(\b r) \cdot [\bhat p, \eps_{2}(\b r)]\big) + \big(\eps_{1} (\b r) \cdot [\bhat p, \eps_{2}(\b r)]\big) \cdot \bhat p \notag \\
		&\phantom{=\frac{\ii}{2 \hbar}(}- \bhat p \cdot \big(\eps_{2}(\b r)\cdot [\bhat p, \eps_{1}(\b r)]\big) - \big(\eps_{2}(\b r) \cdot [\bhat p, \eps_{1}(\b r)]\big) \cdot \bhat p\big] \notag \\
		&= \frac{1}{2} [\bhat p \cdot \eps_\Delta(\b r) + \eps_\Delta(\b r) \cdot \bhat p], \label{eq_deriv_eps_delt_cal}
	\end{align}
	with
	\begin{align} 
		\eps_\Delta(\b r) &= \frac{\ii}{\hbar}\big(\eps_1(\b r) \cdot [\bhat p, \eps_2(\b r)] - \eps_2(\b r) \cdot [\bhat p, \eps_1(\b r)]\big) \notag\\
		&= \eps_1(\b r) \cdot \nabla \eps_2(\b r) - \eps_2(\b r) \cdot \nabla \eps_1(\b r). \label{eq_eps_delta}
	\end{align}
	Note that for smooth vector fields $\eps_1(\b r)$ and $\eps_2(\b r)$, the `difference' vector field $\eps_{\Delta}(\b r)$ is again smooth. {We recall the illustration of $\eps(\rv)$ in Fig.~\ref{FIG1}.}

	In summary, we conclude that
	\begin{align}
		[\Sigma[\eps_1], \Sigma[\eps_2]] &= -\frac{\ii}{\hbar} \Big[\ccdot, \Big[\frac{\ii}{\hbar} \hat \Jcal[\eps_1], \hat \Jcal[\eps_2]\Big]\Big] \notag \\
		&= -\frac{\ii}{\hbar} [\ccdot, \hat \Jcal[\eps_\Delta]] = \Sigma[\eps_\Delta], \label{eq_comm_Sigma}
	\end{align}
	which proves the commutator relation~\eqref{EQSigmaLieAlgebra}. The sketched proof of Eq.~\eqref{EQSigmaLieAlgebra} is based on the Jacobi identity, the linearity and the (Leibniz) product rule of the commutator and the `generalized' canonical commutation relations. Note that the Poisson bracket~\eqref{EQPoissonBracket} satisfies similar properties, so the derivation of the classical version of Eq.~\eqref{EQSigmaLieAlgebra}, see Ref.~\cite{mueller2024gauge}, is similar.
	
	\section{Commutator of shifting superoperators}
	\label{APPcommutatorShiftingOperators}
	
	The commutator of two shifting superoperators~\eqref{EQqsigCommutator} can be derived from the commutator~\eqref{EQSigmaLieAlgebra} of the integrated shifting superoperators. Using Eq.~\eqref{EQlongQuantumSigmaDefinition1}, the commutators are linked via 
	\begin{align}
		[\Sigma[\eps_1], \Sigma[\eps_2]] = \int d \b r \int d \b r^\prime \eps_1(\b r) \cdot [\bg \sigma(\b r), \bg \sigma(\b r^\prime)] \cdot \eps_2(\b r^\prime). \label{eq_Sigma_comm_int_eps}
	\end{align}
	The left-hand side of Eq.~\eqref{eq_Sigma_comm_int_eps} equals the integrated shifting superoperator $\Sigma[\eps_\Delta]$ with shifting field $\eps_\Delta(\b r)$, see Eqs.~\eqref{EQSigmaLieAlgebra} and~\eqref{EQepsDelta}, so exploiting the explicit form of $\eps_\Delta(\b r)$ leads to
	\begin{align}
		&\Sigma[\eps_\Delta] = \int d \b r \bg \sigma(\b r) \cdot \eps_\Delta(\b r) \notag \\
		&= \int d \b r \eps_{1}(\b r) \cdot \big(\nabla \eps_{2}(\b r)\big) \cdot \bg \sigma(\b r) \notag\\
		&\phantom{=} - \int d \b r^\prime \bg \sigma(\b r^\prime) \cdot [\eps_{2}(\b r^\prime) \cdot \big(\nabla^\prime \eps_{1}(\b r^\prime)\big)] \notag\\
		&= \int d \b r \int d \b r^\prime \eps_{1}(\b r) \cdot [\nabla \delta(\b r - \b r^\prime)] \bg \sigma(\b r) \cdot \eps_2(\b r^\prime) \notag\\
		&\phantom{=} - \int d \b r \int d \b r^\prime \eps_1(\b r) \cdot \bg \sigma(\b r^\prime) [\nabla^\prime \delta(\b r - \b r^\prime)] \cdot \eps_2(\b r^\prime) \label{eq_Sigma_eps_delta_int_eps}
	\end{align}
	with $\nabla^\prime = \partial/\partial \b r^\prime$.

	Building the second derivative of Eqs.~\eqref{eq_Sigma_comm_int_eps} and~\eqref{eq_Sigma_eps_delta_int_eps} with respect to $\eps_1(\b r)$ and $\eps_2(\b r)$ together with Eq.~\eqref{EQSigmaLieAlgebra} finally yields the commutator of shifting superoperators 
	\begin{align}
		[\bg \sigma(\b r), \bg \sigma(\b r^\prime)] = [\nabla \delta(\b r - \b r^\prime)] \bg \sigma(\b r) + \bg \sigma(\b r^\prime) [\nabla \delta(\b r - \b r^\prime)], \label{eq_sigma_comm}
	\end{align}
	where we use that $\nabla^\prime \delta(\b r - \b r^\prime) = -\nabla \delta(\b r - \b r^\prime)$, so this proves Eq.~\eqref{EQqsigCommutator}.

	\section{The hyperforce and hypercurrent sum rule for non-Hermitian operators}
	\label{APPnonHermitian}
	
	The hyperforce density balance~\eqref{EQhyperForceDensityBalance} can be generalized for non-Hermitian operators. Such Hilbert space operators $\hat A$ that are not necessarily self-adjoint satisfy
	\begin{align}
		\langle \bhat S_A^\dagger(\b r) \rangle = (\bhat S_A(\b r) \vert 1) = (1 \vert \bhat S_A(\b r))^\star = \langle \bhat S_A(\b r)\rangle^\star, \label{eq_hyp_force_average_dagger}
	\end{align}
	where we have first re-written the thermal average by using the Mori product~\eqref{EQlongQuantumMoriProductDefinintion1}, used its conjugate symmetry~\eqref{eq_mori_conj_symm} and re-identified the thermal average. The same argumentation steps as in Ref.~\cite{mueller2025quantum} lead to the identity $\b S_A^\star(\b r) = - \langle \hat A \bg \sigma(\b r)\rangle  = - (\hat A \vert \beta \bhat F_0(\b r))$. The conjugate symmetry~\eqref{eq_mori_conj_symm} of the Mori product yields the general hyperforce sum rule for non-Hermitian operators: 
	\begin{align} \label{eq_hyp_force_bal_non_herm}
		\b S_A(\b r) + (\beta \bhat F_0(\b r)\vert \hat A) = 0.
	\end{align}
	
	Since the Mori product of Hermitian operators is real, as discussed in Appendix~\ref{APPinnerProducts}, the conjugate symmetry~\eqref{eq_mori_conj_symm} of the Mori product implies for Hermitian operators $\hat A$ the identity $(\hat A \vert \beta \bhat F_0(\b r)) = (\beta \bhat F_0(\b r)\vert \hat A)$. Thus the general hyperforce sum rule~\eqref{eq_hyp_force_bal_non_herm} for non-Hermitian operators is indeed a generalization of Eq.~\eqref{EQhyperForceDensityBalance}.
	
	The same steps (combined with the argumentation in Sec.~\ref{SECdynamicalGaugeInvariance}) show that the hypercurrent sum rule for non-Hermitian operators has the form 
	\begin{align} \label{eq_hyp_curr_bal_non_herm}
		\b S_A(\b r, t) + (\bhat C(\b r, t) \vert \hat A(t)) = 0,
	\end{align}
	which is a generalization of the hypercurrent sum rule~\eqref{EQqsigHypercurrentSumRule}.
	
	\section{The quantum canonical transformation and the shifting superoperator}
	\label{APPquantumCanonical}
	
	The quantum canonical transformation, described in Ref.~\cite{hermann2022quantum} and reproduced in Eqs.~\eqref{EQlongQuantumCanonicalTransformation1} and~\eqref{EQlongQuantumCanonicalTransformation2}, displaces the fundamental degrees of freedom $\b r_i$ and $\bhat p_i$ while the transformed degrees of freedom, denoted by $\bilde r_i$ and $\bildehat p_i$, still satisfy the canonical commutation relations. Here we analyze the connection of the quantum canonical transformation with the shifting superoperator. 
	
	According to Eqs.~\eqref{EQlongQuantumCanonicalTransformationSigma1} and~\eqref{EQlongQuantumCanonicalTransformationSigma2}, the integrated shifting superoperator and the linearized quantum transformation are related since $(1 + \Sigma[\eps]) \b r_i$ equals $\bilde r_i$ and on the other hand $(1 + \Sigma[\eps]) \bhat p_i$ equals $\bildehat p_i$ up to linear order in $\eps(\b r)$. Since $\delta \Sigma[\eps]/\delta \eps(\b r) = \bg \sigma(\b r)$, functional differentiation with respect to $\eps(\b r)$ yields
	\begin{align}
		&\frac{\delta}{\delta \eps(\b r)}\Big\vert_{\eps = 0} \bilde r_i = \bg \sigma(\b r) \b r_i, \label{eq_quant_can_trafo_shift_oper_r}\\
		&\frac{\delta}{\delta \eps(\b r)}\Big\vert_{\eps = 0} \bildehat p_i = \bg \sigma(\b r) \bhat p_i, \label{eq_quant_can_trafo_shift_oper_p}
	\end{align}
where the dyadic structure is indicated by the order of the respective vectors.

	Alternatively, both sides of Eqs.~\eqref{eq_quant_can_trafo_shift_oper_r} and~\eqref{eq_quant_can_trafo_shift_oper_p} can be calculated separately to prove the respective equality, see Ref.~\cite{hermann2022quantum} for more details about the functional derivatives of the transformed degrees of freedom.
	
	Using Eq.~\eqref{eq_quant_can_trafo_shift_oper_r} we observe for a general (smooth) function $\varphi(\b r^N)$ that
	\begin{align}
		\frac{\delta \varphi(\tilde {\b r}^N)}{\delta \eps(\b r)} \Big\vert_{\eps = 0} &= \sum_j \big(\nabla_j \varphi(\b r^N)\big) \cdot \bg \sigma(\b r) \b r_j \notag \\
		&= \bg \sigma(\b r) \varphi(\b r^N), \label{eq_quant_can_trafo_shift_oper_func_r}
	\end{align}
	where we have used the functional chain rule and replaced the functional derivative of $\bilde r_j$ according to Eq.~\eqref{eq_quant_can_trafo_shift_oper_r} in the first step and then used the chain rule in the second step. Note that the shifting superoperator $\bg \sigma(\b r)$ applied to the multiplicative operator $\varphi(\b r^N)$ yields the multiplicative operator $\bhat S_{\varphi}(\b r)$ (see Eq.~\eqref{EQlongQuantumSAOperatorPositions} with $\hat A(\b r^N)$ replaced by $\varphi(\b r^N)$), so the effect of the shifting superoperator on multiplicative operators is analogous to the effect of a differential operator.
	
	Furthermore, we observe for $\gamma=1,..,d$ that
	\begin{align}
		\frac{\delta (\tilde {\hat p}_{i_\gamma})^n}{\delta \eps(\b r)}\Big \vert_{\eps = 0} &= \sum_{k = 1 }^{ n } \hat p_{i_\gamma}^{k-1} [\bg \sigma(\b r) \hat p_{i_\gamma}] \hat p_{i_\gamma}^{n-k+1} \notag \\
		&= \bg \sigma(\b r) \hat p_{i_\gamma}^n \label{eq_quant_can_trafo_shift_oper_poly_p}
	\end{align}
	where we have used the product rule of functional differentiation and replaced the functional derivative of $\bildehat p_i$ according to Eq.~\eqref{eq_quant_can_trafo_shift_oper_p} in the first step and used the product rule of the shifting superoperator $\bg \sigma(\b r)$ in the second step, see Sec.~\ref{SEChigherSumRules}.
	
	In order to connect the quantum canonical transformation and the shifting superoperator with each other, we consider observables $\hat A(\b r^N, \bhat p^N)$ of polynomial structure in $\bhat p^N$ and general dependence on $\b r^N$. { Such operators can be rewritten in the following ordered form:
	\begin{align}
		\hat A(\b r^N, \bhat p^N) = \sum_{\b k_1,..., \b k_n} \varphi_{\b k_1,...,\b k_n}(\b r^N) \bhat p_1^{\b k_1}...\bhat p_N^{\b k_N}, \label{eq_obs_ordered_func}
	\end{align}	
	with $\bhat p_i^{\b k_i} = \hat p_{i_1}^{k_{i_1}} ... \hat p_{i_d}^{k_{i_d}}$, where we sum over all $d$-dimensional vectors $\b k_i = (k_{i_1},...,k_{i_d})$ with (non-negative) integer entries.}
	Then due to Eqs.~\eqref{eq_quant_can_trafo_shift_oper_func_r} and~\eqref{eq_quant_can_trafo_shift_oper_poly_p}, the action of the quantum canonical transformation and that of the shifting superoperator $\bg \sigma(\b r)$, when applied to observables $\hat A(\b r^N, \bhat p^N)$, are linked via
	\begin{align}
		&\frac{\delta \hat A(\bilde r^N, \bildehat p^N)}{\delta \eps(\b r)}\Big\vert_{\eps(\b r) = 0} = \bg \sigma(\b r) \hat A(\b r^N, \bhat p^N), \label{eq_quant_can_trafo_shift_oper_func}
	\end{align}
	with $\hat A(\bilde r^N, \bildehat p^N)$ denoting the observable where $\b r^N$ and $\bhat p^N$ in Eq.~\eqref{eq_obs_ordered_func} are replaced with $\bilde r^N$ and $\bildehat p^N$.
	
	\section{Single-particle averages from many-particle averages}
	\label{APPoccupationNumber}
	
	As is pertinent for Sec.~\ref{SECexchangeSymmetryAndShifting}, choosing either symmetry or anti-symmetry of the Hilbert space basis with respect to particle exchange encodes whether we consider bosons or fermions \cite{bogoliubov1967book, zagoskin2014book, fabrizio2022book}. For additive Hamiltonians, we can construct suitable bases from single-particle eigenfunctions and reduce the many-particle averages to single-particle averages.
	
	More precisely, we consider a Hamiltonian of the additive form $\Hhat = \sum_i \Hcal_i$ where the single-particle Hamiltonians $\Hcal_i$ act on the variable $\b r_i$ via $\Hcal_i \Psi(\b r_1, ..., \b r_N) = \Hcal \Psi(\b r_1,..., \b r_{i-1}, \b r, \b r_{i+1}, ..., \b r_N) \vert_{\b r = \b r_i}$ with Hamiltonian $\Hcal$ acting on single-particle functions $\varphi(\b r)$. Furthermore, the single-particle Hamiltonian $\Hcal$ induces an orthonormal basis $\varphi_n(\b r)$ of eigenfuctions, i.e. $\Hcal \varphi_n(\b r) = e_n \varphi_n(\b r)$ with eigenvalue $e_n$.
	
	The orthonormal basis $\varphi_n(\b r)$ induces an anti-symmetric and a symmetric many-particle orthonormal basis. The anti-symmetric basis is given by
	\begin{align} \label{eq_anti_symm_basis}
		\psi_{\b n}(\b r^N) &= \frac{1}{(N!)^{\frac{1}{2}}} \sum_{\kappa \in S_N} \sgn(\kappa) \varphi_{n_1}(\b r_{\kappa(1)}) ...\varphi_{n_N}(\b r_{\kappa(N)})\\
		&= (N!)^{-\frac{1}{2}} \det\big(\varphi_{n_i}(\b r_j)\big),
	\end{align}
	for $\b n=n_1,...,n_N$ with $n_1 < ... < n_N$, which is the Slater determinant, and the symmetric basis is given by
	\begin{align} \label{eq_symm_basis}
		\psi_{\b n}(\b r^N) = \frac{1}{(\alpha_1! ... \alpha_k! N!)^{\frac{1}{2}}} \sum_{\kappa \in S_N} \varphi_{n_1}(\b r_{\kappa(1)}) ...\varphi_{n_N}(\b r_{\kappa(N)}),
	\end{align}
	for $\b n = n_1,...,n_N$ with $n_1 \leq ... \leq n_N$, where the index $n_1,...,n_N$ consists of $k$ different indices and $\alpha_j$ denotes the number of appearance in $n_1,...,n_N$, so that $\alpha_1 + ... + \alpha_k = N$. Here $S_N$ denotes the set of all permutation on $\{1,..,N\}$ which has cardinality $N!$ and $\sgn(\kappa)$ is the sign of the permutation $\kappa \in S_N$. We have used the same symbol $\psi_{\bf n}(\b r^N)$ to denote the different bases~\eqref{eq_anti_symm_basis} and~\eqref{eq_symm_basis}.
	
	Note that anti-symmetric or symmetric many-particle functions $\Psi(\b r^N)$ satisfy
	\begin{align}
		\Psi(\b r_1, ..., \b r_N) = s(\kappa) \Psi(\b r_{\kappa(1)}, ..., \b r_{\kappa(N)})
	\end{align}
	for all permutations $\kappa \in S_N$ with $s(\kappa) = \sgn(\kappa)$ for anti-symmetry and $s(\kappa) = 1$ for symmetry. This is inline with the definition given in Sec.~\ref{SECexchangeSymmetryAndShifting}, since the exchange of two indices is a permutation with sign $-1$ and every permutation can be constructed by elementary transpositions.
	
	Since both bases consist of single-particle eigenfunctions of $\Hcal$, the additive structure of the Hamiltonian $\Hhat = \sum_i \Hcal_i$ implies that both the anti-symmetric~\eqref{eq_anti_symm_basis} and the symmetric~\eqref{eq_symm_basis} basis functions $\psi_{\b n}(\b r^N)$ are eigenfunctions of the Hamiltonian $\Hhat$ with eigenvalues $E_{\b n} = \sum_{i} e_{n_i}$ for $n_1 \leq ... \leq n_N$ in the symmetric case and for $n_1 < ... <n_N$ in the anti-symmetric case.
	
	Using the anti-symmetric basis~\eqref{eq_anti_symm_basis} or the symmetric basis~\eqref{eq_symm_basis}, the many-particle thermal average of an observable $\hat A$ of the same additive structure $\hat A = \sum_i \Acal_i$ as the Hamiltonian with associated single-particle observable $\Acal$ simplifies to the weighted sum over single-particle thermal averages
	\begin{align}
		\langle \hat A \rangle &= \sum_{\b n} \Big \langle \psi_{\b n} \Big| \sum_i \Acal_i \frac{\e^{-\beta \Hhat}}{Z} \Big| \psi_{\b n} \Big \rangle \notag\\
		&=\sum_{\b n} \frac{\e^{-\beta E_{\b n}}}{Z} \sum_i \langle \varphi_{n_i}(\b r) | \Acal | \varphi_{n_i}(\b r) \rangle,
	\end{align}
	where $\sum_{\b n}$ denotes the sum over all indices with $n_1 \leq ... \leq n_N$ or $n_1 < ... < n_N$, depending on whether the chosen basis is symmetric or anti-symmetric.

	\section{Compatibility with particle index permutations}
	\label{APPexchange}
	
	Similar to the classical shifting operator~\cite{mueller2024gauge}, see Eq.~\eqref{EQbsigClassicalPoisson}, the quantum shifting superoperator $\bg \sigma(\b r)$ is also invariant under particle index permutation, as discussed in Sec.~\ref{SECexchangeSymmetryAndShifting}. In particular, the shifting superoperator respects symmetries of functions under particle index exchange. Such symmetries are for example encoded already within the chosen the basis functions, see Appendix~\ref{APPoccupationNumber} for (anti-)symmetric orthonormal basis functions constructed from a single-particle orthonormal basis.
	
	We consider a general permutation $\kappa: \{1,...,N\} \to \{1,...,N\}$ with permutation operator $\Pcal_\kappa$ given by $\Pcal_\kappa f(\b r^N) = f(\b r_{\kappa(1)},...,\b r_{\kappa(N)})$. Note that the exchange of two particle indices, described in Sec.~\ref{SECexchangeSymmetryAndShifting}, is one special special class of permutation.
	
	Straightforward calculations show that
	\begin{align} \label{eq_perm_deriv}
		\Pcal_\kappa \nabla_j = \nabla_{\kappa(j)} \Pcal_\kappa,
	\end{align} 
	so $\bhat p_i = - \ii \hbar \nabla_i$ implies $\Pcal_\kappa \bhat p_j = \bhat p_{\kappa(j)} \Pcal_\kappa$. Furthermore using that $\Pcal_\kappa g(\b r_j) = g(\b r_{\kappa(j)})$, we conclude $\Pcal_\kappa \delta(\b r - \b r_j) = \delta(\b r - \b r_{\kappa(j)}) \Pcal_\kappa$.
	
	Exploiting these relations, we obtain 
	\begin{align}
		&\Pcal_\kappa \sum_{i} [\bhat p_i \delta(\b r - \b r_i) + \delta(\b r - \b r_i) \bhat p_i] \notag\\
		&= \sum_{i} [\bhat p_{\kappa(i)} \delta(\b r - \b r_{\kappa(i)}) + \delta(\b r - \b r_{\kappa(i)}) \bhat p_{\kappa(i)}] \Pcal_\kappa. \label{eq_perm_sum_current}
	\end{align}
	Since the order of summation over all particles is irrelevant and the permutation $\kappa$ is bijective, the sum of the left-hand side and of the right-hand side of Eq.~\eqref{eq_perm_sum_current} are equal, so we conclude
	\begin{align}
		\Pcal_\kappa m \bhat J(\b r) = m \bhat J(\b r) \Pcal_\kappa. \label{eq_perm_current}
	\end{align}
	In particular, Eq.~\eqref{eq_perm_current} implies $\Pcal_\kappa [\hat A, m \bhat J(\b r)] = [\Pcal_\kappa \hat A, m\bhat J(\b r)]$. This shows the compatibility of the shifting superoperator $\bg \sigma(\b r)$ and particle index permutation,
	\begin{align} \label{eq_perm_sigma}
		\Pcal_\kappa \bg \sigma(\b r) \hat A = \bg \sigma(\b r) \Pcal_\kappa \hat A,
	\end{align}
	so $[\Pcal_\kappa, \bg \sigma(\b r)] = 0$, where $\Pcal_\kappa$ is interpreted as a superoperator.
	
	\section{Thermal averages in the extended average}
	\label{APPextendendEnsemble}
	
	We consider the thermal average of the extended ensemble, introduced in Sec.~\ref{SECextendedEnsemble}. For convenience, we recall that the modified Hamiltonian~\eqref{EQextendedHamiltonian} has the form $\Hhat_{A}=\Hhat_0 -\lambda \hat A/\beta$ and thermal averages are built via $\langle \ccdot \rangle_A = \Tr \ccdot \e^{-\beta \Hhat_A}/Z_A$ with extended partition sum $Z_A = \Tr \e^{-\beta \Hhat_A}$. Thus the partial derivative of the thermal average in the extended ensemble with respect to $\lambda$ has the form
	\begin{align} \label{eq_aver_ext_ens_deriv}
		\frac{\partial }{\partial \lambda}\langle \hat B \rangle_A = \Big\langle \frac{\partial \hat B}{\partial \lambda} \Big\rangle_A + \cov_A(\hat A \vert \hat B),
	\end{align}
	where $(\hat C \vert \hat B)_A$ and $\cov_A(\hat C \vert \hat B) = (\hat C \vert \hat B)_A - \langle \hat B \rangle_A \langle \hat C^\dagger \rangle_A$ are the Mori product~\eqref{EQlongQuantumMoriProductDefinintion1} and the Mori covariance~\eqref{EQlongQuantumMoriCovariance} with Hamiltonian $\Hhat_A$. 
	
	To prove Eq.~\eqref{eq_aver_ext_ens_deriv} we calculate the partial derivative on the left-hand side of Eq.~\eqref{eq_aver_ext_ens_deriv} explicitly. Using the product rule leads to the following three terms
	\begin{align}
		&\frac{\partial}{\partial \lambda} \langle \hat B \rangle_A = \frac{\partial}{\partial \lambda} \Tr \hat B \e^{-\beta \Hhat_A}/Z_A \\
		&= \Tr \Big[\frac{\partial}{\partial \lambda} \hat B \Big] \e^{-\beta \Hhat_A}/Z_A + \Tr \hat B \Big[\frac{\partial}{\partial \lambda} \e^{-\beta \Hhat_A}\Big]/Z_A \notag \\
		&\phantom{=} + \Tr \hat B \e^{-\beta \Hhat_A} \Big[\frac{\partial}{\partial \lambda}\frac{1}{Z_A}\Big], \label{eq_aver_ext_ens_deriv_three_term}
	\end{align}
	where the first term in Eq.~\eqref{eq_aver_ext_ens_deriv_three_term} equals the first term in Eq.~\eqref{eq_aver_ext_ens_deriv} due to the definition of the thermal average $\langle \ccdot\rangle_A$ in the extended ensemble.
	
	The crucial step to make progress with the second and third terms in Eq.~\eqref{eq_aver_ext_ens_deriv_three_term} is to calculate $\partial \e^{-\beta \Hhat_A}/\partial \lambda$ which can be written as $\partial \e^{-\beta \Hhat_A}/\partial \lambda = [\partial/\partial \lambda, \e^{-\beta \Hhat_A}]$. We use the commutator relation~\eqref{eq_comm_exp_func} for exponentiated operators, discussed in Appendix~\ref{APPcommutatorExp} below, and obtain that
	\begin{align} 
		\Big[\frac{\partial}{\partial \lambda}, \e^{-\beta \Hhat_A}\Big] &= - \int_{0}^{\beta} d \beta^\prime \e^{\beta^\prime \Hhat_A} \Big[\frac{\partial}{\partial \lambda}, \Hhat_A\Big] \e^{\beta^\prime \Hhat_A} \e^{-\beta \Hhat_A} \notag \\
		&= \beta^{-1} \int_{0}^{\beta} d \beta^\prime \e^{\beta^\prime \Hhat_A} \hat A \e^{\beta^\prime \Hhat_A} \e^{-\beta \Hhat_A}, \label{eq_comm_exp_func_appl_deriv_lamb}
	\end{align}
	where we exploit that $[\partial/\partial \lambda, \Hhat_A] = \partial \Hhat_A/\partial \lambda = -\hat A/\beta$, as follows from the form~\eqref{EQextendedHamiltonian} of the extended Hamiltonian.
	
	Relation~\eqref{eq_comm_exp_func_appl_deriv_lamb} shows that the second term of Eq.~\eqref{eq_aver_ext_ens_deriv_three_term} is linked to the Mori product via
	\begin{align}
		&\Tr \hat B \Big(\frac{\partial}{\partial \lambda} \e^{-\beta \Hhat_A}\Big)\Big/ Z_A \notag \\
		&= \beta^{-1} \int_{0}^{\beta} d \beta^\prime \hat B\e^{-\beta^\prime \Hhat_A} \hat A \e^{\beta^\prime \Hhat_A} \e^{-\beta \Hhat_A}/Z_A \notag \\
		&= (\hat B^\dagger \vert \hat A)_A = (\hat A \vert \hat B)_A, \label{eq_aver_ext_ens_deriv_sec_term}
	\end{align}
	where we use that the self-adjointness of $\hat A$ together with Eq.~\eqref{eq_mori_adj} implies $(\hat B^\dagger \vert \hat A)_A = (\hat B^\dagger \vert \hat A^\dagger)_A = (\hat A \vert \hat B)_A$. 
	
	Furthermore, relation~\eqref{eq_comm_exp_func_appl_deriv_lamb} simplifies the partial derivative of the partition sum via
	\begin{align}
		\frac{\partial Z_A}{\partial \lambda} & = \Tr \Big(\frac{\partial}{\partial \lambda} \e^{-\beta \Hhat_A}\Big) \notag \\
		& = \beta^{-1} \int_{0}^{\beta} d \beta^\prime \Tr \e^{\beta^\prime \Hhat_A} \hat A \e^{\beta^\prime \Hhat_A} \e^{-\beta \Hhat_A} \notag \\
		&= \Tr \hat A \e^{-\beta \Hhat_A} = Z_A\langle \hat A\rangle_A, \label{eq_aver_ext_ens_deriv_part_sum}
	\end{align}
	where we exploit additionally the invariance of the quantum trace under cyclic permutations, see Appendix~\ref{APPinvarianceTrace}, and the definition of the extended thermal average $\langle \ccdot \rangle_A$.
	
	Since $\partial Z_A^{-1}/\partial \lambda = - Z_A^{-2} \partial Z_A/\partial \lambda = - \langle \hat A\rangle_A/Z_A$, the third term of Eq.~\eqref{eq_aver_ext_ens_deriv_three_term} simplifies to
	\begin{align}
		\Tr \hat B \e^{-\beta \Hhat_A} \Big(\frac{\partial}{\partial \lambda}\frac{1}{Z_A}\Big)	&= -\Tr \hat B \e^{-\beta \Hhat_A}/Z_A \langle \hat A \rangle_A \notag \\
		&= -\langle \hat B \rangle_A \langle \hat A \rangle_A, \label{eq_aver_ext_ens_deriv_third_term}
	\end{align}
	where we have identified again the extended thermal average in the last step.
	
	In summary, we conclude that
	\begin{align}
		\frac{\partial}{\partial \lambda} \langle \hat B \rangle_A &= \Big \langle \frac{\partial \hat B}{\partial \lambda} \Big \rangle_A + (\hat A \vert \hat B)_A - \langle \hat B \rangle_A \langle \hat A \rangle_A, \label{eq_aver_ext_ens_deriv_summary}
	\end{align}
	which is identical to Eq.~\eqref{eq_aver_ext_ens_deriv} due to the definition~\eqref{EQlongQuantumMoriCovariance} of the Mori covariance with Hamiltonian $\Hhat_A$.
	
	Note that the argumentation remains valid when replacing the canonical trace $\Tr \cdot$ with the grand-canonical trace $\Tr^\prime \,\cdot\,=
\sum_{N=0}^\infty\sum_n \langle n|\,\cdot\,|n\rangle e^{\beta\mu
  N}/N!$, the Boltzmann factor $\e^{-\beta \Hhat_A}$ with $\e^{-\beta (\Hhat_A - \mu N)}$, and the partition sum $Z_A$ with the grand partition sum $\Xi = \Tr^\prime \e^{-\beta (\Hhat_A - \mu N)}$. Hence the above results hold analogously in the grand canonical ensemble.

	\section{Commutator of an exponentiated operator}
	\label{APPcommutatorExp}
	
	An important expression (see e.g.\ Appendix~\ref{APPextendendEnsemble}) is the commutator of a Hilbert space operator $\hat B$ and an exponentiated operator $\e^{-\beta \hat C}$. This can be written as \cite{kubo1957}:
 	\begin{align}
		[\hat B, \e^{-\beta \hat C}] = -\int_0^\beta d\beta'\e^{-\beta'\hat C}[\hat B, \hat C]\e^{\beta' \hat C}\e^{-\beta\hat C}. \label{eq_comm_exp_func}
	\end{align}
	The identity~\eqref{eq_comm_exp_func} is used in Ref.~\cite{mueller2025quantum} to derive the hyperforce sum rule~\eqref{EQforceDensityBalance} and it can be proven by applying the operator $[\hat B, \e^{-\beta \hat C}] \e^{\beta \hat C}$ to an arbitrary function $f(\b r^N)$. Differentiation with respect to $\beta$ yields
	\begin{align}
		\frac{\partial}{\partial \beta}  \big([\hat B, \e^{-\beta \hat C}] \e^{\beta \hat C} f(\b r^N)\big) = - \e^{-\beta \hat C} [\hat B, \hat C] \e^{\beta \hat C}f(\b r^N). \label{eq_comm_exp_func_deriv}
	\end{align}
	Integrating this identity with respect to $\beta$ leads to
	\begin{align}
		[\hat B, \e^{-\beta \hat C}] \e^{\beta \hat C} f(\b r^N) = - \int_{0}^{\beta} d \beta^\prime \e^{-\beta^\prime \hat C} [\hat B, \hat C] \e^{\beta^\prime \hat C}f(\b r^N), \label{eq_comm_exp_func_int}
	\end{align}
	where we use that $[\hat B, \e^{-\beta \hat C}] \e^{\beta \hat C} f(\b r^N)\vert_{\beta = 0} = 0$.
	By replacing $f(\b r^N)$ with $\e^{-\beta \hat C} f(\b r^N)$ and leaving away $f(\rv^N)$ on both sides, we obtain the exponential commutator rule~\eqref{eq_comm_exp_func}.
	
	\section{Kubo identity and hypercurrent sum rule}
	\label{APPrelationshipKuboIdentity}
	
	The hypercurrent balance~\eqref{EQqsigHypercurrentSumRule} can be written in an alternative form. Replacing in Eq.~\eqref{eq_quant_trace_prod} $\hat A$ with $(\ii/\hslash) m \bhat J(\b r, t)$, $\hat B $ with $\e^{-\beta \Hhat_0}$ and $\hat C$ with $\hat A(t)$, the anti-linearity of the commutator yields the identity
	\begin{align}
		\Tr \Big[\frac{\ii}{\hbar} m\bhat J(\b r, t), \hat A(t)\Big] \e^{- \beta \Hhat_0} = \Tr \frac{\ii}{\hbar} m\bhat J(\b r,t) [\hat A(t),\e^{- \beta \Hhat_0}].
	\end{align}
	Similar argumentation steps as in Sec.~\ref{SECdynamicalGaugeInvariance} yield the following alternative form of the hypercurrent balance~\eqref{EQqsigHypercurrentSumRule}:
	\begin{align}
		\b S_A(\b r,t) &= \Tr \frac{\ii}{\hbar}[m\bhat J(\b r, t), \hat A(t)] \e^{- \beta \Hhat_0}/Z \notag \\
		&= \Tr m\bhat J(\b r,t) \frac{\ii}{\hbar}[\hat A(t), \e^{- \beta \Hhat_0}/Z] \notag \\
		&= \Big(m\bhat J(\b r,t) \Big\vert \frac{\ii}{\hbar}[\hat A(t), -\beta \Hhat_0]\Big). \label{eq_hyp_curr_kubo_id}
	\end{align}
	
	Equation~\eqref{eq_hyp_curr_kubo_id} is a special case of the Kubo identity presented in Ref.~\cite{fick1990book}. Specifically, the Kubo identity~\cite{fick1990book} has the form 
	\begin{align} \label{eq_kubo_id}
		(\hat B \vert {\sf L} \hat C) = \frac{1}{\beta \hbar} \langle [\hat B^\dagger, \hat C]\rangle
	\end{align}
	with Hilbert space operators $\hat B$ and $\hat C$ and Liouville superoperator ${\mathsf L} = (1/\hbar) [\Hhat_0, \ccdot]$.
	Replacing $\hat B$ and $\hat C$ with $m \bhat J(\b r,t)$ and $(\ii/\hbar) \hat A(t)$, the definition of the shifting superoperator~\eqref{EQqsigDefinition} shows that Eq.~\eqref{eq_hyp_curr_kubo_id} is indeed a special case of the generic Kubo identity~\eqref{eq_kubo_id}.
The present gauge correlation framework goes beyond the more generic Green-Kubo formalism by revealing the specific Lie algebra structure \eqref{EQSigmaLieAlgebra} that characterizes the shifting superoperator. This leads to exact identities such as the commutator sum rule \eqref{EQcommutatorSumRule}.

\end{document}